\documentclass[12pt,a4paper]{article} 
\usepackage{amssymb}
\def \ni {\noindent} 
\def \vs {\vskip5mm}

\def \mea {\nonumber\\}

\def \AutP {{\rm Aut}({\bf P})} 
\def \AutV {{\rm Aut}({\bf V})} 
\def \AO {{\hat A}} 
\def \RO {{{\hat \rho}\,\,}} 
\def \calN {{\cal N}} 
\def \calZ {{\cal Z}} 
\def \UO {{\hat U}} 
\def \VO {{\hat V}} 
\def \qO {{\hat q}} 
\def \pO {{\hat p}} 
 
\def \Hspace {{\cal H}} 
\def \Kspace {{\cal K}} 
\def \AF {A(q,p)} 
 
\def \Wmap {{\cal W}} 
\def \Winv {{\cal W}^{-1}} 
\def \Tspace {{\cal T}} 
\def \eO {{\hat e}} 
\def \AO {{\hat A}} 
\def \RO {{\hat \rho}} 
\def \UO {{\hat U}} 
\def \VO {{\hat V}} 
\def \XO {{\hat X}} 
\def \CO {{\hat C}} 
\def \BO {{\hat B}} 
\def \HO {{\hat H}} 
\def \Jspace {{\cal J}} 
\def \Jdual {{\cal J'}} 
\def \Sspace {{\cal S}} 
\def \Gspace {{\cal G}} 
\def \Qspace {{\cal Q}} 
\def \Gdual {{\cal G'}} 
\def \Sdual {{\cal S'}} 
\def \be {\begin{equation}} 
\def \ee {\end{equation}} 
\def \bea {\begin{eqnarray}} 
\def \eea {\end{eqnarray}} 
\begin{document} 
\title{Quantum Symmetries and the Weyl-Wigner Product of Group Representations}
\author{A.J. Bracken\cite{AJBaddress}$^{1,2}$, G. Cassinelli$^{1}$ and J.G. Wood$^{2}$\\
$\,^{1}$DIFI, Universit\`a di Genova\\
Via Dodecaneso 33\\
Genova, 16146 ITALY\\
and\\
$\,^{2}$Centre for Mathematical Physics\\  Department of Mathematics\\
The University of Queensland\\
Brisbane,  4072
AUSTRALIA}
\maketitle
\begin{abstract}
In the usual formulation of quantum mechanics, 
groups of automorphisms of quantum states have ray representations by unitary and antiunitary
operators on complex Hilbert space, in accordance with
Wigner's Theorem.  In the phase-space formulation, they have real, true unitary representations
in the space of square-integrable functions on phase-space. 
Each such phase-space representation is a Weyl-Wigner product of the corresponding Hilbert space
representation with its contragredient, and these can be recovered by `factorising' the
Weyl-Wigner product.  However, 
not every real, unitary representation on phase-space corresponds to a
group of automorphisms,  so not every such representation is in the form of a Weyl-Wigner
product and can be factorised.  The conditions under which this is possible are examined.
Examples are presented.    

\end{abstract}
\section{Introduction} 
Since the pioneering works of Weyl \cite{weyl}, von Neumann \cite{vonneumann}, 
Wigner \cite{wigner},
Groenewold \cite{groenewold}  and Moyal \cite{moyal}, the phase space
formulation of quantum mechanics has been the subject of much research from many different points
of view.  The    
underlying theory has been greatly 
developed
\cite{takabayasi}-\cite{sternheimer}, including 
group-theoretical aspects 
\cite{segal}-\cite{dragt}
of particular relevance to the present work.
\vs\ni
Our interest here is in the way that 
symmetries, and more generally, groups of automorphisms of quantum states, are expressed  
by group representations in the formulation of quantum mechanics on phase space $\Gamma$, 
and the relationship of these  to 
more familiar representations on complex Hilbert space $\Hspace$.   
Representations of automorphism groups on $\Gamma$ are typically true, 
real, unitary representations, whereas 
representations on $\Hspace$ can be projective,  
and even antiunitary  
(such as in the case of time-reversal symmetry), in accordance with  
Wigner's Theorem \cite{wigner2,bargmann,cassinelli}. 
Such a phase space representation $\Pi_{\Gamma}$ is isomorphic  (but not equal) \cite{arnal} to
the tensor product of a corresponding
Hilbert space representation 
$\Pi_{\Hspace}$
with its contragredient 
$\Pi_{\Hspace}^C$.  
We call it the Weyl-Wigner product of 
$\Pi_{\Hspace}$
and
$\Pi_{\Hspace}^C$, and write
\be
\Pi_{\Gamma}=
\Pi_{\Hspace}\stackrel{W}{\otimes}\Pi_{\Hspace}^C
\cong
\Pi_{\Hspace}\otimes\Pi_{\Hspace}^C\,.
\label{weylwigproddef}
\ee
\vs\ni
Recent successes of `quantum tomography' \cite{leonhardt} have highlighted the fact that the
quantum state vector (wavefunction) in $\Hspace$ can be recovered  
from the Wigner distribution function on $\Gamma$, up to a constant phase \cite{tatarskii}.
In principle,  the whole  
Hilbert space structure of quantum mechanics
can be recovered from the phase space structure \cite{dubin}, so we must expect 
that a 
projective, complex, unitary or antiunitary representation $\Pi_{\Hspace}$ in Hilbert space 
can 
be recovered  
from the corresponding true, real, 
unitary representation $\Pi_{\Gamma}$  in phase space, in effect
by 
`factorising' $\Pi_{\Gamma}$ as a 
Weyl-Wigner product (\ref{weylwigproddef}).  We shall confirm that this is 
the case. It is 
remarkable that this is possible, in  particular
because ray representations are  
associated with central extensions at the Lie algebra level, and it
can only happen if the  
associated extension parameters 
(mass of a particle, Planck's constant, ...)
already appear in the true, phase space representation, or else 
arise in the mapping from phase space
back to Hilbert space.  We shall see that both possibilities are realized.  
\vs\ni 
The structure of the phase space formulation in its original form is intimately connected 
with the structure of the Heisenberg-Weyl group. Extensions  
to other groups have been described \cite{stratonovich,agarwal2,wolf}, but 
we shall deal here only with
the original form, restricting $\Gamma$ to the phase plane  coordinatised by  
the pair $(q,p)$.
However, we shall be concerned with representations on $\Gamma$ and $\Hspace$
of groups and Lie algebras other than  
the Heisenberg-Weyl group and algebra.  
Generalizations to quantum systems with several degrees of freedom, and systems with spin,
are certainly possible.

\vs\ni 
At the heart of the phase space formulation of 
quantum mechanics lies the Weyl-Wigner transform $\Wmap$,
which is an invertible mapping  
from linear operators $\AO$, $\BO$, $\dots$ 
on
$\Hspace$ to functions  
$A$, $B$, $\dots$ on $\Gamma$\,.  
\vs\ni 
Before embarking on a  discussion of automorphism groups and their 
representations, it is necessary to  
outline a firm mathematical basis for 
the Weyl-Wigner transform and its inverse. 
More detail can be found in the literature  \cite{pool,dubin,sternheimer}.  
We work with dimensionless 
variables in what follows, in effect setting Planck's constant $\hbar$
equal to 1, except in the first two examples at the end of the paper.

\section{ Background: A mathematical setting for the  
Weyl-Wigner transform} 
\vs\ni 
For our purposes, an appropriate 
setting for a description of  
$\Wmap$ and $\Wmap ^{-1}$ for a system with one degree of freedom involves 
\cite{segal,pool,dubin}
\begin{itemize} 
\item[ $\bullet$] 
the complex vector space of Hilbert-Schmidt operators on $\Hspace$, regarded as a Hilbert space 
$\Tspace _C$ with scalar product 
\be 
(\AO,\BO)_{\Tspace_C}={\rm Tr}\,(\AO^{\dagger}\BO)\,, 
\label{Tspacescalarprod} 
\ee 
\item[ $\bullet$] 
the complex vector space $L_2\,({\mathbb C},d\Gamma)$, regarded as a Hilbert space  
$\Kspace_C$ with scalar product 
\be 
(A,B)_{\Kspace_C}=\frac{1}{2\pi}
\int {\overline A}\, B\,d\Gamma\,,\quad d\Gamma=dq\,dp\,, 
\label{Kspacescalarprod} 
\ee 
\end{itemize} 
together with certain associated vector spaces. 
(We use ${\rm Tr}$ to denote the trace, and the overbar to denote complex conjugation. 
Integrals are over all real values of the variables of 
integration, unless otherwise indicated.) 
 
\vs\ni 
The Hilbert space $\Hspace$ of state vectors can be realised  
as $L_2\,({\mathbb C},dx)$ (the `coordinate representation') 
with scalar product  
\be 
(\varphi,\psi)_{\Hspace} =\int \overline{\varphi(x)}\psi(x)\,dx\,. 
\label{Hspacescalarprod} 
\ee  
\vs\ni 
Let $e_1\,,\,e_2\,,\,\dots$ form an orthonormal basis of `test' functions 
in this realisation of $\Hspace$. Each $e_r$ and its Fourier transform 
is infinitely differentiable and each, together with all its  derivatives,  
vanishes more quickly than any negative power of its 
argument, as that argument approaches $\pm\infty$;  
the eigenfunctions of the Hamiltonian operator of a simple harmonic oscillator 
provide an example. 
Introduce the `Gel'fand triple' of vector spaces 
 
\be 
\Gspace <\Hspace <\Gdual\,. 
\label{Htripledef} 
\ee 
where $\Gspace$ is the Schwartz space associated with the basis  
$\{e_r\}$,  
and $\Gdual$ is its strong dual 
\cite{gelfand,roberts,antoine,bohm,dubin}.

\vs\ni 
Now let $\eO_{rs}$, for $r,\,s=1,\,2,\,\dots \,$ denote the 
rank-1 operator on $\Hspace$ corresponding to the above choice of basis, defined by 
\be 
\eO_{rs}\varphi = (e_s,\varphi)_{\Hspace}\,\,e_r\,,\quad\forall\, \varphi\in\Hspace\,. 
\label{rankoneopdef} 
\ee 
Then the set of $\eO_{rs}$ forms an orthonormal basis in $\Tspace_C$, with 
\be 
(\eO_{rs},\eO_{uv})_{\Tspace_C}=\delta_{ru}\delta_{sv}\,. 
\label{Tbasis} 
\ee 
Introduce the Gel'fand triple  
\be 
\Sspace_C <\Tspace_C <\Sdual_C\,. 
\label{Ttripledef} 
\ee 
by analogy with (\ref{Htripledef}).  
 
\vs\ni 
Corresponding to each $\eO_{rs}\in\Tspace_C$, define $\Phi_{rs}\in\Kspace_C$, by 
\be 
\Phi_{rs}(q,p)= 
\int 
e_r(q-y/2)\overline{e_s(q+y/2)} 
\,e^{ipy}\,dy\,. 
\label{Phidef} 
\ee 
It is easily checked that the set of $\Phi_{rs}$ forms an 
orthonormal basis in $\Kspace_C$, and that each $\Phi_{rs}$ is a `test function' 
of two variables.  
Introduce the Gel'fand triple 
\be 
\Jspace_C<\Kspace_C<\Jdual_C 
\label{Ktripledef} 
\ee 
by analogy with (\ref{Htripledef}) and (\ref{Ttripledef}).  
 
\vs\ni 
The elements of $\Gdual$ are `generalised functions' on the real line. 
Similarly, $\Jdual_C$ consists of 
generalised functions on the phase plane. 
The elements of $\Sdual_C$ are `generalised linear operators,'  
and include the operators in $\Sspace_C$ and $\Tspace_C$.  
It is not difficult to see that a generalised linear operator 
can be regarded as carrying 
elements of $\Gspace$ into elements of $\Gdual$ in general,  
that is to say, test functions of one variable into generalised functions of one variable
\cite{dubin}.  
\vs\ni 
The Weyl-Wigner transform is a 1-1 invertible mapping from $\Sdual_C$ onto $\Jdual_C$  
which associates a generalised function $A$ with each  
generalised operator $\AO$. 
Note firstly that each $\AO\in\Sspace_C$ 
can be interpreted as an integral operator  
\be 
(\AO\varphi)(x)=\int A_K(x,y)\varphi(y)\,dy\,,\quad \varphi\in\Hspace\,, 
\label{intopA} 
\ee 
whose kernel $A_K$ is a test function of two variables. Then  
define 
\be 
A(q,p)=(\Wmap(\AO))(q,p)=\int A_K(q-y/2,q+y/2)\,e^{ipy}\,dy\,, 
\label{WmapA} 
\ee  
with inverse 
\be 
A_K(x,y)=(\Winv (A))_K(x,y)= \frac{1}{2\pi}\int A((x+y)/2,p)\,e^{ip(x-y)}\,dp\,. 
\label{WinvmapA} 
\ee 
These formulas (\ref{WmapA}) and (\ref{WinvmapA}) can be 
extended to apply to every operator in $\Tspace_C$ (regarded as an integral operator) 
and every function in $\Kspace_C$ if the integrals are interpreted in the usual 
generalised way for Fourier transforms of $L_2$ functions. 
\vs\ni 
Once $\Wmap$ and $\Winv$ have been defined in this way
on $\Sspace_C$ and $\Jspace_C$, respectively,  
their definitions can be extended easily to $\Sdual_C$ and $\Jdual_C$, 
respectively, as follows.  
For each $ 
\tau 
\in \Sdual _C$, define  
$ 
\Wmap( 
\tau) 
\in \Jdual_C$ by  
\be 
\Wmap(\tau)(\kappa) =\tau(\Wmap ^{-1} (\kappa))\,, \forall\, \kappa\in \Jspace_C\,, 
\label{extendWdef1} 
\ee 
\vs\ni 
and conversely, for each $ 
\kappa\in\Jdual_C 
$, define  
$ 
\Wmap^{-1}( 
\kappa)\in\Sdual_C$  
by 
\be 
\Wmap^{-1}(\kappa)(\tau)= \kappa(\Wmap (\tau))\,,\forall\, \tau\in \Sspace_C\,. 
\label{extendWdef2} 
\ee 
This defines $\Wmap$ and $\Wmap ^{-1}$ as mappings from $\Sdual_C$ onto $\Jdual_C$ and 
{\it vice versa}.  The mappings are continuous in the natural topologies 
on these spaces \cite{dubin}.  
  
\vs\ni 
In particular, $\Wmap$ and $\Wmap ^{-1}$ map $\Tspace_C$ onto $\Kspace_C$ and {\it vice versa}.  
In this case, as can be seen from (\ref{WmapA}) and (\ref{WinvmapA}),  
we have for every $\AO$, $\BO\in \Tspace_C$ and corresponding 
$A$, $B\in\Kspace_C$,  
  
\be 
(A,B)_{\Kspace_C}=(\AO,\BO)_{\Tspace _C} 
\,, 
\label{unitarity} 
\ee 
showing that $\Wmap$ and $\Winv$ act as unitary transformations from  
$\Tspace_C$ onto $\Kspace _C$ 
and  
{\em vice versa}. 
\vs\ni 
We note that  
$\Sdual_C$ contains two important classes 
of operators with the property that 
every operator in each class
has every $\psi\in\Gspace$ 
in its domain\,: 
\begin{itemize} 
\item[ $\bullet$] 
The class of Hilbert-Schmidt operators, forming $\Tspace_C$,  
which are bounded and defined on all of $\Hspace$.  
\item[ $\bullet$] 
The class $Q$ of operators which leave $\Gspace$ invariant, and so have $\Gspace$ 
as a common, invariant domain, dense in $\Hspace$.  
This class contains in particular 
the unit operator ${\hat I}$ on $\Hspace$ and the canonical operators $\qO$, $\pO$ defined  
on $\psi\in\Gspace$ by 
\be 
\qO\psi(x)=x\psi(x)\,,\quad \pO\psi(x) = -i \psi'(x)\,, 
\label{canonicalops} 
\ee  
and it therefore also contains all polynomials in these operators, forming a subclass 
$Q_{WH}\subset Q$. We can say that 
$Q_{WH}$ defines a representation on $\Gspace$ of the enveloping algebra of the Heisenberg-Weyl 
Lie algebra.  
\end{itemize} 
 
\vs\ni 
The classes $\Tspace_C$, $Q$ and $Q_{WH}$ share another important property\,:  
each is invariant under the formation of operator products.  
Note that $\Tspace_C$ and $Q$ are not disjoint, and that neither is a subclass 
of the other.  
\vs\ni 
For $\AO$, $\BO\in\Tspace_C$, we define 
the associative but noncommutative
star product \cite{groenewold,moyal} of the corresponding $A$, $B\in\Kspace_C$ 
by  
\be 
A\star B 
\,\,(\,\,= \Wmap(\AO)\star\Wmap(\BO)\,\,) 
\,\,=\Wmap(\AO\BO)\,. 
\label{stardefA} 
\ee 
The Wigner transform defines not only a 
unitary transformation from $\Tspace_C$ to $\Kspace_C$, but also 
an isomorphism 
of these two sets, regarded as algebras.  
The usual operator product in $\Tspace_C$ 
is replaced by the 
star product of functions in $\Kspace_C$. 
The image of  
($-i\times$) the commutator on $\Tspace_C$ is the Groenewold-Moyal \cite{groenewold,moyal}
bracket on $\Kspace_C\,$: 
\be 
\{A,B\}_{GM}=-i(A\star B-B\star A)\,. 
\label{moyalA} 
\ee 
For sufficiently smooth $A$ and $B$, in particular for $A$, $B\in\Jspace_C$, it can be  
seen from (\ref{WmapA}) and (\ref{WinvmapA}) that  
\bea 
(A\star B)(q_1,p_1)\qquad\qquad\qquad\qquad\qquad\qquad 
\qquad 
\qquad 
\qquad 
\qquad 
\qquad 
\mea 
=\frac{1}{\pi^2} 
\int 
A(q_2,p_2)B(q_3,p_3)e^{-2i[p_1(q_2-q_3)+p_2(q_3-q_1)+p_3(q_1-q_2)]}\,dq_2dp_2dq_3dp_3\,, 
\label{starprod2} 
\eea 
and so  
\bea 
\{A,B\}_{GM} (q_1,p_1)= 
- \frac{2}{\pi^2} 
\int 
A(q_2,p_2)B(q_3,p_3) 
\qquad 
\qquad 
\qquad 
\qquad 
\qquad 
\mea 
\times \sin(2[p_1(q_2-q_3)+p_2(q_3-q_1)+p_3(q_1-q_2)])\,dq_2dp_2dq_3dp_3\,. 
\label{moyal2} 
\eea 
For such $A$ and $B$, the order of the integrations is 
unimportant. For general $A$, $B\in\Kspace_C$, (\ref{starprod2}) and (\ref{moyal2}) are valid 
with a generalised interpretation of the integrals.

\vs\ni 
The image under $\Wmap$ of the space $Q_{WH}$ is the subspace ${\cal I}_{WH}$ of $\Jdual_C$, 
consisting of polynomials in $1$, $q$ and $p$.  
In particular, $\Wmap({\hat I})=1$, $\Wmap(\qO)=q$ and $\Wmap(\pO)=p$.  
For $\AO$, $\BO\in Q_{WH}$, we again use (\ref{stardefA}) and (\ref{moyalA}) to 
define the star product and Groenewold-Moyal 
bracket of the corresponding $A$, $B\in{\cal I}_{WH}$. 
The transforms $\Wmap$ and $\Winv$ map $\Qspace_{WH}$ onto ${\cal I}_{WH}$ and  
{\em vice versa}, preserving polynomial degree. 
This action establishes an equivalence of two representations of the enveloping algebra 
of the Heisenberg-Weyl Lie 
algebra, one in $\Qspace_{WH}$ with the usual operator product, 
the other in ${\cal I}_{WH}$ with the star product. The structure of the 
mappings $\Wmap$ and $\Wmap^{-1}$ in this case is  
well known \cite{weyl,agarwal,berezin,tatarskii}.  
We have 
\be 
\Wmap(\qO ^{\,n} \,\pO ^{\, m})=  
\sum_{k=0}^{{\rm min}\,(m,n)} \left(\frac{i}{2}
\right)^k k!\,C^m_k \,C^n_k \,q^{\,m-k}\,p^{\,n-k}\,,  
\label{berezin1} 
\ee 
where $C^m_r=m!/(r!(m-r)!)$. 
Conversely,  
\bea 
\Wmap^{-1} 
(q^m\,p^n)&=& 
\sum_{k=0}^{{\rm min}\,(m,n)} \left(\frac{-i}{2}\right)^k k!
\,C^m_k\, C^n_k \,\qO^{\,m-k}\,\pO^{\,n-k}\,,
\mea 
&=&\frac{1}{2^m}\sum_{r=0}^m C^m_r\, \qO^{\,m-r}\pO^{\,n}\qO^{\,r}\,. 
\label{berezin2} 
\eea 
Further similar formulas can be obtained by replacing 
$\qO$ by $\pO$, $\pO$ by $-\qO$, $q$ by $p$, and $p$ by $-q$.  
\vs\ni 
On ${\cal I}_{WH}$, 
the star product reduces to 
\bea 
(A\star B)(q,p) &=& 
A(q,p) B(q,p)+i (A\,J\, B)(q,p) - \frac{1}{2!}(A\,J^2\, B)(q,p)+ \dots  
\mea &=&  
B(q,p) A(q,p)- i (B\,J\, A)(q,p) - \frac{1}{2!}(B\,J^2 \,A)(q,p)+ \dots \,, 
\mea 
\mea 
\label{starprodWH} 
\eea 
where  
\be 
J = 
\frac{1}{2}(\frac{\partial ^{(L)}}{\partial q}\frac{\partial ^{(R)}}{\partial p} 
-\frac{\partial ^{(R)}}{\partial q}\frac{\partial ^{(L)}}{\partial p})\,, 
\label{janus1} 
\ee 
with $L$ and $R$ indicating the directions in which the 
differential operators act. Then the Groenewold-Moyal bracket (\ref{moyalA}) 
takes the form 
\bea 
&&(\{A,B\}_{GM}) (q,p)= 
2\big((A\,J\, B)(q,p)  
-\frac{1}{3!}(A\,J^3 \,B)(q,p) 
\mea
&&\qquad+\frac{1}{5!}(A\,J^5\, B)(q,p) 
+ \dots 
\label{moyalB} 
\eea 
The formulas (\ref{starprodWH}) and (\ref{moyalB}) are commonly written as 
\bea 
A\star B=A e^{iJ}B=Be^{-iJ}A 
\mea 
\{A,B\}_{GM}=2A\sin(J)B\,. 
\label{expstarandmoyal} 
\eea 
Note however that because $A$ and $B$ in $Q_{WH}$ are polynomials, 
the series in (\ref{starprodWH}) and (\ref{moyalB}) terminate.  
\vs\ni 
More generally, we can use (\ref{stardefA}) and (\ref{moyalA}) 
to define the star product and Groenewold-Moyal bracket  
of those $A$, $B\in\Jdual_C$ corresponding to 
$\AO$, $\BO\in Q$. Note that, for all $\AO\in\Sdual_C$, 
\be 
\Wmap(\AO)=A\Leftrightarrow \Wmap(\AO^{\dagger})=\overline{A}\,, 
\label{Aconj} 
\ee 
and that whenever the star product of  
$A$ and $B$ is defined, it satisfies 
\be 
\overline{A\star B}=\overline{B}\star\overline{A}\,, 
\label{starconj} 
\ee 
reflecting the fact that 
\be 
(\AO\BO)^{\dagger}=\BO^{\dagger}\AO^{\dagger}\,. 
\label{opconj} 
\ee

\section{Quantum states}  
\ni 
Let $\Tspace_R$ denote  
the Hilbert space of self-adjoint Hilbert-Schmidt operators 
over the real numbers, with scalar product 
\be 
(\AO,\BO)_{\Tspace_R}={\rm Tr}\,(\AO\BO)\,. 
\label{TspaceRscalarprod} 
\ee 
Its image under $\Wmap$ is $\Kspace_R$, the Hilbert space of 
square-integrable, real-valued functions on $\Gamma$ 
with scalar product 
\be 
(A,B)_{\Kspace_R} = 
\frac{1}{2\pi} 
\int A\,B\,d\Gamma\,. 
\label{KspaceRscalarprod} 
\ee 
The elements of $\Tspace_R$ and $\Kspace_R$ 
represent a class of observables on a quantum system, 
in the Hilbert space and phase space formulations, respectively.  
By an obvious extension of the arguments for $\Tspace_C$ and 
$\Kspace_C$, the mappings $\Wmap$ and $\Winv$ act as unitary transformations between 
$\Tspace_R$ and $\Kspace_R$\,: if $A=\Wmap(\AO)$ and $B=\Wmap(\BO)$, we have  
\be 
(\AO,\BO)_{\Tspace_R}=(A,B)_{\Kspace_R}\,. 
\label{realunitarity} 
\ee

\vs\ni 
Let  
${\bf P}\subset\Tspace_R$  
denote the set  
of pure state density operators  
$\RO\in 
\Tspace _R$ for the quantum system, which are characterised by the conditions  
\be 
\RO ^2 =\RO\,,\quad (\RO,\RO)_{\Tspace _R} =1\,. 
\label{purerhos} 
\ee 
The set of pure and mixed state density operators is  
the convex set of $\RO\in\Tspace _R$ with the 
pure state density operators as extremal points.  
Corresponding to each $\RO$, pure or mixed, the Wigner distribution function is defined as 
\be 
W=\frac{1}{2\pi} \,\Wmap (\RO)\,.  
\label{Wdef1} 
\ee 
It follows at once from (\ref{TspaceRscalarprod}), 
(\ref{KspaceRscalarprod}) and (\ref{realunitarity}) that 
\be 
{\rm Tr}\,(\RO\AO)=\int W(q,p) \AF \,d\Gamma\,, 
\label{means} 
\ee 
for each $\AO\in\Tspace_R$ and corresponding $A\in\Kspace_R$,  
which equates the familiar expressions for quantum averages  
in the Hilbert space and phase space forms.

\vs\ni 
Let ${\bf V}=\Wmap({\bf P})\subset \Kspace_R$ denote the set of pure state Wigner functions. 
For any $W\in{\bf V}$ we have, from(\ref{Wdef1}) and (\ref{purerhos}),
\be
W\star W = \frac{1}{2\pi} W\,,\quad 2\pi \int W^2 
\,d\Gamma =\int W\,d\Gamma =1\,.
\label{purewigs}
\ee
Pure state density operators, and hence pure state Wigner functions, 
are in one-to-one correspondence with unit rays in the Hilbert space of state vectors.

\vs\ni 
Given a unit ray, 
the corresponding pure state density operator 
is the 1-dimensional projection whose action on any $\chi\in\Hspace$ is given by 
\be 
\RO\chi=(\psi,\chi)_{\Hspace}\,\,\psi\,, 
\label{rhoaction} 
\ee 
where $\psi$ is any vector in  the ray.  In the coordinate representation adopted 
in Section 2, $\RO$ is the integral operator with kernel 
$\psi(x)\overline{\psi(y)}$, and the corresponding Wigner function 
takes the form, from (\ref{WmapA}),
\be 
W(q,p)= 
\frac{1}{2\pi} 
\int 
\psi(q-y/2) 
\overline{\psi(q+y/2)} 
e^{ipy} 
\,dy\,. 
\label{wigfndef} 
\ee

\vs\ni 
The inverse problem, of finding  
the unit ray
corresponding to a given  
pure state Wigner function $W$ satisfying (\ref{purewigs}),  
which is equivalent to the problem of finding  
the unit ray corresponding to a given pure state density operator $\RO$  
satisfying (\ref{purerhos}), 
has been treated  
by Tatarskii \cite{tatarskii}.  However, it is difficult to define a linear mapping
from  Wigner functions to  corresponding wavefunctions.  This is an obstacle to recovering a
(linear) unitary mapping between wavefunctions in $\Hspace$, corresponding to a given mapping
between Wigner functions, associated with a transformation from some symmetry group, say. 
On the other hand, it is known that the Hilbert space structure is represented within the phase
space structure \cite{dubin}.   We shall see that it is possible in principle
to recover such unitary symmetry
operators
on $\Hspace$, as well as antiunitary symmetry operators, using a different approach.

\section{Automorphisms and Wigner's Theorem} 
\ni 
Let  
Aut(${\bf P}$)  
denote the set of automorphisms of 
${\bf P}$.  
It consists of all bijective maps 
$\mu \,:\,{\bf P}\to {\bf P}$ that also satisfy the condition 
\be 
(\mu (\RO _1),\mu (\RO_2))_{\Tspace_R} = (\RO_1,\RO_2)_{\Tspace _R} 
\label{autrho1} 
\ee 
for all $\RO_1$, $\RO_2$ $\in {\bf P}$, and 
is a group under the natural composition of mappings.  
We refer to the mappings in Aut(${\bf P}$) as ${\bf P}$-{\em automorphisms}.  
\vs\ni 
Let  
Aut(${\bf V}$) denote the set of automorphisms of ${\bf V}$.  
It  consists of all bijective maps 
$M\,:\,{\bf V}\to {\bf V}$ that also satisfy the condition 
\be 
(M(W_1),M(W_2))_{\Kspace_R} = (W_1,W_2)_{\Kspace _R} 
\label{autW1} 
\ee 
for all $W_1$, $W_2$ $\in {\bf V}$, and similarly forms a group. 
We refer to the mappings in Aut(${\bf V}$) as ${\bf V}$-{\em automorphisms}.  
\vs\ni 
The Weyl-Wigner transform defines a unitary transformation from $\Tspace_R$ to $\Kspace _R$ 
which maps ${\bf P}$ onto ${\bf V}$, and establishes an isomorphism of Aut(${\bf P}$) and  
Aut(${\bf V}$). Explicitly, 
\be 
M(W)= M(\Wmap ({\hat \rho}))=\Wmap(\mu({\hat \rho}))\,,
\quad \mu({\hat \rho})=\mu (\Winv (W))=\Winv (M(W))\,. 
\label{autaction1} 
\ee  
\vs\ni 
According to Wigner's Theorem \cite{wigner2,bargmann,cassinelli}, 
given any $\mu\in$ Aut(${\bf P}$), there 
exists a unitary or antiunitary operator ${\hat U}$ on $\Hspace$, 
unique up to a phase factor, such that 
\be 
\mu ( {\hat \rho})= {\hat U} {\hat \rho}{\hat U}^{\dagger}
\quad \forall\, {\hat \rho}\in {\bf P}\,. 
\label{rhoautaction} 
\ee 
 
\vs\ni 
Each $\mu\in$ Aut(${\bf P}$) extends to an operator on $\Tspace_C$, with 
\be 
\mu (\AO)=\UO \AO \UO^{\dagger}\,,\quad\forall \AO\in\Tspace_C\,. 
\label{genmuaction} 
\ee 
This operator does not act linearly on $\Tspace_C$ in general, 
but it does always act linearly on $\Tspace_R$, 
which it leaves invariant. It defines a real unitary 
transformation of $\Tspace_R$ onto itself. We denote this transformation also by $\mu$.
We extend the whole group of automorphisms $\AutP$ in this way to act on all of $\Tspace_R$,
and denote this group with extended domain of action also by $\AutP$. 
Similarly, we extend each automorphism  $M\in\AutV$, and hence $\AutV$ itself, to act unitarily
on all of
$\Kspace_R$.  

\vs\ni
Aut(${\bf P}$)  
is isomorphic to $\Sigma (\Hspace)$,  
the group of unitary and antiunitary operators on $\Hspace$, 
factored by its closed centre, the phase-group \cite{cassinelli}\,: 
\be 
\Sigma (\Hspace)={\bf U}\cup{\bf {\overline U}}/{\bf T}\,. 
\label{symmetrygroup} 
\ee 
and it follows that this is also true of  
Aut(${\bf V}$).  
\vs\ni 
In the case that $\UO$ is unitary,  
the action of $M$ on $A\in \Kspace_R$ corresponding to (\ref{genmuaction}) is given by  
\be 
M(A)=\Wmap({\hat U}\AO{\hat U}^{\dagger})=
\Wmap({\hat U})\star A\star \Wmap({\hat U}^{\dagger}) 
=U\star A\star {\overline U}\,. 
\label{Wautaction1} 
\ee 
Here $U=\Wmap(\UO)$ is a complex-valued function on $\Gamma$, 
and $\overline{U}$ is its complex conjugate. Corresponding to unitarity of $\UO$ 
we have 
\be 
U\star \overline{U}=\overline{U}\star U =1\,. 
\label{starunitarity} 
\ee 
Note 
that a unitary operator $\UO$  
lies in $\Sdual_C$,  
not in $\Tspace_C$, but 
\be 
\AO\in\Tspace_R\Rightarrow \UO\AO\UO^{\dagger}\in\Tspace_R\,. 
\label{UAUdagger} 
\ee 
Similarly, a star-unitary function  
$U$ lies in $\Jdual_C$, not $\Kspace_C$, but 
\be A\in\Kspace_R \Rightarrow U\star A\star \overline{U}\in\Kspace_R\,.  
\label{UAUbar} 
\ee 
\vs\ni 
In the case that $\UO$ is antiunitary, because the action of 
$\Wmap$ on antiunitary operators has not been defined, 
we proceed as follows.
Consider the particular antiunitary 
operator $\CO$ on $\Hspace$ which leaves all basis vectors 
$e_r$ invariant: if $\varphi=\sum_r \varphi_r e_r$, then  
\be 
\CO\varphi=\sum_r \overline{\varphi_r} e_r\,. 
\label{Caction} 
\ee  
(If we work in the coordinate representation, 
and choose a basis in $\Hspace$ of real-valued functions,  
then $\CO$ is the operation of complex conjugation.)  
 
\vs\ni 
Next, let ${\cal P}$ denote the operator on $\Kspace_C$ defined by 
\be 
({\cal P}(A))(q,p)=A(q,-p)\quad {\rm for}\quad {\rm all}\quad A\in \Kspace_C\,. 
\label{Pdef} 
\ee  
Then ${\cal P}$ is  unitary on $\Kspace_C$, and also (real) unitary when restricted to 
$\Kspace_R$.  It is evident that, on $\Kspace_C$ or $\Kspace_R$, 
\be 
{\cal P}^{\dagger}={\cal P}\,,\quad {\cal P}^2=I\,. 
\label{Pprops} 
\ee 
Direct calculation from (\ref{starprod2}) shows also that 
\be 
{\cal P}(A\star B)={\cal P}(B)\star {\cal P}(A)\,\,\,{\rm for} 
\,\,\,{\rm all}\,\,\, A,\,B\in 
\Kspace_C\,.  
\label{Pidentity} 
\ee 
 
\vs\ni 
The form of the Wigner function  
$W'$ corresponding to $\RO'=\CO\RO \CO$ is now given from (\ref{wigfndef}) by  
\be 
W'(q,p)=W(q,-p)=({\cal P}(W))(q,p)\,, 
\label{Pdef2}
\ee 
and generalising (\ref{Pdef2}), 
we find for a general $\AO\in\Tspace_C$ and corresponding $A\in\Kspace_C$, that  
\be 
\Wmap (\CO \AO\CO)={\cal P}(\overline{A})\,. 
\label{genPaction1} 
\ee 
The transformation of $W$ corresponding to a general antiunitary 
operator $\UO= \CO\VO $ in (\ref{rhoautaction}), where $\VO$  
is unitary, is  
\bea 
W'={\cal P}(V\star W\star \overline{V})={\cal P}(\overline{V})\star 
{\cal P}(W)\star {\cal P}(V)\,,\quad{\rm or} 
\mea 
W'(q,p)=(V\star W\star \overline{V})(q,-p) \,,  
\label{genPaction2} 
\eea 
where $V=\Wmap(\VO)$ satisfies the star-unitarity condition (\ref{starunitarity}).  
More generally, for any $\AO\in\Tspace_C$ and corresponding $A\in\Kspace_C$,
\be
\Wmap(\UO\AO\UO^{\dagger})=\Wmap(\CO\VO\AO\VO^{\dagger}\CO) 
={\cal P}(V\star\overline{A}\star \overline{V})={\cal P} 
(\overline{V})\star{\cal P}(\overline{A}) 
\star{\cal P}(V)\,.
\label{genPaction3}
\ee
\vs\ni 
Let  
$O(\Tspace_R)$  
denote the group of all real unitary transformations of $\Tspace_R$ onto itself.
and let $O(\Kspace_R)$  
denote the group 
of real unitary transformations of  
$\Kspace_R$ onto itself. Then  
\bea 
\AutP\cong \AutV\,,\quad O(\Tspace_R)\cong O(\Kspace_R) 
\mea 
\AutP<O(\Tspace_R)\,,\quad  
\AutV<O(\Kspace_R)\,. 
\label{subgroup1} 
\eea 
In particular, it is important to note that in general
$\AutP$ and $\AutV$ are  proper subgroups of $O(\Tspace_R)$ and $O(\Kspace_R)$, respectively.
It is easy to see that $\AutP$ can be  characterised as the subgroup  
of $O(\Tspace_R)$ whose elements satisfy
\be
\mu(\AO\BO)=\mu(\AO)\mu(\BO)\quad\forall \AO,\,\BO\in\Tspace_R\,.
\label{AutPprop}
\ee
For if $\mu\in\AutP$, then (\ref{AutPprop}) is satisfied as a consequence of (\ref{genmuaction}),
and conversely, if $\mu\in O(\Tspace_R)$ satisfies (\ref{AutPprop}), then it 
is a bijective map from 
${\bf P}$ 
to ${\bf P}$ 
which satisfies (\ref{autrho1}),
and so belongs to $\AutP$. 
Likewise $\AutV$ can be characterised as the subgroup of $O(\Kspace_R)$ whose elements satisfy
\be
M(A\star B)=M(A)\star M(B)\,.
\label{AutVprop}
\ee

\vs\ni 
Given an element $\mu\in$ Aut(${\bf P}$) 
(or equivalently, given an element $M\in\AutV$), 
it is possible in principle to  construct
the corresponding unitary or antiunitary operator 
$\UO$ of (\ref{genmuaction}), up to a phase, and  
proofs of Wigner's Theorem show how it can be done \cite{wigner2,bargmann,cassinelli}. 
However, there   
seems to be no simple recipe for such a construction  in general. Fortunately,  
in many applications to physics, 
we have to deal with connected Lie groups of automorphisms,  
possibly extended by discrete transformations, and the problem
of identifying 
the  generator of a 1-parameter
group of unitaries corresponding to the generator of 
a given 1-parameter group of automorphisms is more straightforward.
This is exploited in what follows.

\section{Symmetries and the Weyl-Wigner product}
\ni
Given a group $G$ and a quantum system 
having $\Hspace$ as its space of state vectors, 
we say that $G$ is a {\em pre-symmetry group} of the system,  
if there exists a homomorphism $\mu$ from $G$ onto 
a subgroup  ${\bar G}<\AutP$. 
The group of symmetries of the Hamiltonian of the system 
serves as one example, and any dynamical symmetry (or spectrum-generating) group 
as another,  but $\AutP$ is large, with many subgroups. 
Wigner's 
Theorem \cite{wigner2,bargmann,cassinelli} shows that $\AutP$, and hence
every pre-symmetry group $G$, has
a ray representation
$\Pi_{\Hspace}$ by unitary and antiunitary operators on $\Hspace$, 
\bea
\Pi_{\Hspace}(g)\varphi &=&\UO(g)\varphi\,,  
\mea
\UO(g_1)\UO(g_2)&=&e^{i\omega(g_1,g_2)}\UO(g_1g_2)\,,
\label{Hspacegroupaction}
\eea
where $\omega$ is a real-valued function satisfying appropriate associativity conditions 
\cite{wigner2,bargmann}.
\vs\ni
Of more direct interest to us here is that $G$ has 
a real unitary representation $\Pi_{\Tspace_R}$ on $\Tspace_R$, 
and an isomorphic real unitary representation $\Pi_{\Kspace_R}$ on $\Kspace_R$.  
The representation $\Pi_{\Tspace_R}$ is defined by the action (\ref{genmuaction}) 
of each element $\mu (g)=\Pi_{\Tspace_R}(g)$ 
of ${\bar G}<\AutP$ on an arbitrary element $\AO\in\Tspace_R$\,: 
\be 
g: \, \AO \longrightarrow \Pi_{\Tspace_R}(g)(\AO)=\UO (g)\AO\UO(g)^{\dagger}\,. 
\label{Tspacegroupaction} 
\ee 
The transformation $\Pi_{\Tspace_R}(g)$ is real and unitary, even  
in the case that $\UO (g)$ is antiunitary, 
as noted earlier. 
The group representation property is immediate from (\ref{Tspacegroupaction}): 
\bea 
\Pi_{\Tspace_R}(g_1)\Pi_{\Tspace_R}(g_2)(\AO)
&=& 
\UO(g_1) \UO(g_2) \AO \UO(g_2)^{\dagger}\UO(g_1)^{\dagger} 
\mea 
&=& 
e^{i\omega(g_1,g_2)}\UO(g_1g_2) \AO \UO(g_1g_2)^{\dagger}e^{-i\omega(g_1,g_2)}
\mea 
&=& 
\Pi_{\Tspace_R}(g_1g_2)(\AO)\,,
\label{PiTaction}
\eea

\vs\ni
The unitary representation $\Pi_{\Kspace_R}$ is defined as the  
Weyl-Wigner transform of the  
unitary representation $\Pi_{\Tspace_R}$, to which it is therefore isomorphic:
\bea
\Pi_{\Kspace_R}(g)(A)&=&\Wmap(\Pi_{\Tspace_R}(g)(\AO))\quad{\rm for}\,\,\,{\rm all}\,\,\, 
\AO\in\Tspace_R\,,\,\,\,
\mea
{\rm that}\,\,\,{\rm is}\quad
\Pi_{\Kspace_R}\Wmap&=&\Wmap\Pi_{\Tspace_R}\,\,\,{\rm on}\,\,\, \Tspace_R\,. 
\label{isomorphism2}
\eea

\vs\ni 
The group action of $\Pi_{\Kspace_R}$  follows from that of 
$\Pi_{\Tspace_R}$ in (\ref{PiTaction}), but is worth considering in more detail.
In the case that $\UO (g)$ is unitary,
the action of $\Pi_{\Kspace_R}$
corresponding to (\ref{Tspacegroupaction}) is
\be
g: A\longrightarrow \Pi_{\Kspace_R} (g)(A) 
= U(g)\star A\star \overline{U(g)}\,,  
\label{Kspacegroupaction2} 
\ee 
where 
$U(g)=\Wmap(\UO(g))$ is star-unitary.   
 
\vs\ni 
If  every element $\UO(g)$ 
of $\Pi_{\Hspace}$ is unitary 
then, just as 
\be 
\UO(g_1)\UO(g_2)=e^{i\omega (g_1,g_2)} \UO(g_1g_2)
\label{projectiveaction1} 
\ee 
in (\ref{Hspacegroupaction}), 
so the functions $U(g)$ satisfy  
\be 
U(g_1)\star U(g_2)=e^{i\omega (g_1,g_2)} U(g_1g_2)\,,
\label{projectiveaction2} 
\ee 
and provide a unitary ray representation 
under the star product,  
isomorphic to $\Pi_{\Hspace}$.
Such $\star$-representations  
have been discussed in the literature \cite{segal,kostant,kirillov,fronsdal,arnal,sternheimer}. 
\vs\ni
In the case that $\UO(g)=\CO\VO(g)$ 
is antiunitary, with $\VO(g)$ unitary and $\CO$ the
antiunitary operator in
(\ref{Caction}), the action of $\Pi_{\Kspace_R}(g)$
is, corresponding to (\ref{Tspacegroupaction}), 
\be
g: A\longrightarrow \Pi_{\Kspace_R} (g)(A) = 
{\cal P}\left(V(g)\star A\star \overline{V(g)}\right)= 
{\cal P}\left(\overline{V(g)}\right)\star
 {\cal P}(A) \star {\cal P}(V(g))\,,  
\label{Kspacegroupaction3} 
\ee 
where 
$V(g)=\Wmap(\VO(g))$ is star-unitary.
The group representation property for $\Pi_{\Kspace_R}$, 
which is of course also guaranteed by the isomorphism between 
$\Pi_{\Tspace_R}$ and $\Pi_{\Kspace_R}$,
can be regarded as a consequence of the star-unitarity of 
$\UO(g)$ and $\VO(g)$, and the properties (\ref{Pprops}) of ${\cal P}$.  
For example, if $\UO(g_2)$ is unitary, but $\UO(g_1)=\CO\VO(g_1)$ and 
$\UO(g_1g_2)=\CO\VO(g_1g_2)$ are antiunitary,
then 
\bea
&&\Pi_{\Kspace_R}(g_1)\Pi_{\Kspace_R}(g_2)(A)\\
&&={\cal P}(\overline{V(g_1)})\star{\cal P}(\overline{U(g_2)}) 
\star {\cal P}(A) \star{\cal P}(U(g_2))\star{\cal P}(V(g_1))
\mea
&&= {\cal P}(\overline{U(g_2)}\star\overline{V(g_1)}) 
\star {\cal P}(A)\star{\cal P}(V(g_1)\star U(g_2)) 
\mea 
&&={\cal P}(\overline{(V(g_1)\star U(g_2))} 
\star {\cal P}(A)\star 
{\cal P}(V(g_1)\star U(g_2)) 
\mea 
&&={\cal P}(\overline{V(g_1g_2)})\star {\cal P}(A) \star 
{\cal P}(V(g_1g_2)) 
\mea 
&&={\cal P}(V(g_1g_2)\star A\star\overline{V(g_1g_2)})
\mea
&&= \Pi_{\Kspace_R}(g_1g_2)(A)\,.
\label{Kspacegroupaction4}
\eea

\vs\ni
The representation $\Pi_{\Tspace_R}$ on $\Tspace_R$, 
and hence the representation $\Pi_{\Kspace_R}$ on $\Kspace_R$, is
isomorphic to the tensor product of the Hilbert space representation 
$\Pi_{\Hspace}$ with its contragredient \cite{arnal}: 
\be
\Pi_{\Tspace_R}\cong \Pi_{\Kspace_R}\cong \Pi_{\Hspace}\otimes \Pi_{\Hspace}^C\,.
\label{isomorphisms}
\ee 
To see this, we realize $\Hspace\otimes\Hspace$ as  
$L_2({\mathbb C},dx)\otimes L_2({\mathbb C},dy)$, then $\Pi_{\Hspace}$ on $L_2({\mathbb C},dx)$,  
and $\Pi_{\Hspace}^C$ on 
$L_2({\mathbb C},dy)$. Consider firstly the case that every element of $\Pi_{\Hspace}$ is unitary. 
Let $\calN$ denote the unitary mapping from $\Tspace_R$ to $L_2({\mathbb
C},dx)\otimes L_2({\mathbb C},dy)$ 
defined by  
\be 
\calN (\AO)=A_K\,, 
\label{Naction1} 
\ee 
where  $A_K(x,y)$ is the kernel of $\AO$, regarded as an integral operator, as in (\ref{intopA}). 
Then 
\be 
\calN (\Pi_{\Tspace_R}(g)(\AO))=\int 
U_K(g|x,x')A_K(x',y')\overline{U_K(g|y',y)}\,dx'\,dy'\,, 
\label{PicrossPiCaction1} 
\ee 
corresponding to (\ref{Tspacegroupaction}).  
In (\ref{PicrossPiCaction1}),  
the kernel of $\UO(g)$ is 
$U_K(g|x',y')$, which is not itself square-integrable.
Because the action of $\Pi_{\Hspace}(g)$ in $L_2({\mathbb C},dx)$ is defined by 
\be 
(\Pi_{\Hspace}(g)\varphi)(x)=\int U_K(g|x,x')\varphi (x')\,dx' 
\label{PiHaction} 
\ee 
and the action of $\Pi_{\Hspace}^C(g)$ in $L_2({\mathbb C},dy)$ is defined by 
\be 
(\Pi_{\Hspace}^C(g)\varphi)(y)=\int \overline{U_K(g|y',y)}\varphi (y')\,dy'\,, 
\label{PiHaction2} 
\ee  
then (\ref{PicrossPiCaction1}) expresses the isomorphism between $\Pi_{\Tspace_R}$ and 
$\Pi_{\Hspace}\otimes \Pi_{\Hspace}^C$\,:  
\bea 
\calN(\Pi_{\Tspace_R}(g)(\AO))&=&(\Pi_{L_2({\mathbb C},dx)}(g)\otimes \Pi_{L_2({\mathbb C},dy)}^C(g)) 
(\calN (\AO))\,,\quad{\rm or} 
\mea 
\calN\Pi_{\Tspace_R}&=&(\Pi_{L_2({\mathbb C},dx)}(g)\otimes \Pi_{L_2({\mathbb
C},dy)}^C) 
\calN \quad{\rm on}\,\,\,\Tspace_R\,. 
\label{isomorphism1} 
\eea 
The same is true in the case that $\UO(g)=\CO\VO(g)$ is antiunitary, with $\VO(g)$ 
unitary.  Then 
\be 
\calN (\Pi_{\Tspace_R}(g)(\AO))=\int 
\overline{V_K(g|x,x')}\overline{A_K(x',y')}V_K(g|y',y)\,dx'\,dy'\,, 
\label{PicrossPiCaction2}
\ee 
and because 
\be 
(\Pi_{\Hspace}(g)\varphi)(x)=\int \overline{V_K(g|x,x')}\overline{\varphi (x')}\,dx' 
\label{PiHaction3} 
\ee 
and 
\be 
(\Pi_{\Hspace}^C(g)\varphi)(y)=\int V_K(g|y',y)\overline{\varphi (y')}\,dy'\,, 
\label{PiHaction4} 
\ee 
then (\ref{isomorphism1}) again holds.  
\vs\ni 
Now let $\calZ$ denote the unitary mapping from $L_2({\mathbb C},dx)\otimes L_2({\mathbb C},dy)$ to 
$\Kspace_R$ defined by 
\be 
\calZ=\Wmap\calN^{\dagger}\,. 
\label{Zdef1} 
\ee 
It is not hard to see from (\ref{Naction1}) and (\ref{WmapA}), that  
\be 
(\calZ f)(q,p)=\int f(q-x/2,q+x/2)e^{ipx}\,dx\,, 
\label{Zdef} 
\ee 
with inverse acting as  
\be 
(\calZ^{\dagger}F)(x,y)=\frac{1}{2\pi}\int F((x+y)/2,p)e^{ip(x-y)}\,dp\,. 
\label{Zinvdef} 
\ee 
From (\ref{isomorphism2}) and (\ref{Naction1}),
we have the isomorphism between $\Pi_{\Kspace_R}$ and 
$\Pi_{\Hspace}\otimes\Pi_{\Hspace}^C$ in the form 
\be 
\Pi_{\Kspace_R}\calZ=\calZ(\Pi_{L_2({\mathbb C},dx)}\otimes \Pi_{L_2({\mathbb
C},dy)}^C)\,. 
\label{isomorphism3} 
\ee 
We say that  $\Pi_{\Kspace_R}$ is the Weyl-Wigner product of 
$\Pi_{\Hspace}$ and $\Pi_{\Hspace}^C$, denoted
\be
\Pi_{\Hspace}
\stackrel{W}{\otimes}
\Pi_{\Hspace}^C\,.
\label{wwproddefagain}
\ee

\vs\ni
The reduction to irreducibles of the Weyl-Wigner product 
will evidently lead to the same Clebsch-Gordan series 
as the reduction of the usual tensor product 
$\Pi_{\Hspace}\otimes\Pi_{\Hspace}^C$, 
and the basis vectors on which the reduction is
accomplished will be related by the intertwiner ${\cal Z}$. 
We shall consider this further only in the context of 
Example 4 (Case A) in the next Section, where the
reductions can easily be worked out and compared.

\section{Factorising  phase space representations}
\ni
Not every  real, unitary representation 
$\Pi_{\Kspace_R}$ of a group on the function space $\Kspace_R$
is in the form of a Weyl-Wigner product.  Only those representations forming subgroups of  
$\AutV<O(\Kspace_R)$ have this form.   
In view of (\ref{AutVprop}), the extra  condition to be satisfied is  
\be
\Pi_{\Kspace_R}(g)(A\star B)=\Pi_{\Kspace_R}(g)(A)\star \Pi_{\Kspace_R}(g)(B)  
\label{AutVcondn2}
\ee
for all $A$, $B\in\Kspace_R$ and all $g$ in the group.  
Given a representation $\Pi_{\Kspace_R}$ which is 
in $\AutV$, and so does satisfy (\ref{AutVcondn2}),
it follows from Wigner's Theorem that it 
must be possible to factorise $\Pi_{\Kspace_R}$ as the Weyl-Wigner product of a 
representation $\Pi_{\Hspace}$ on Hilbert space with its contragredient $\Pi_{\Hspace}^C$,
and that this representation $\Pi_{\Hspace}$ will be 
in general a unitary or antiunitary ray representation of the 
underlying group.    
\vs\ni
We now examine how this factorisation process can be put into effect,
and  begin by  specialising to the case of a connected Lie 
group $G$ with a unitary ray representation  
$\Pi_{\Hspace}$  
on  
$\Hspace$,  
a corresponding real unitary representation 
$\Pi_{\Tspace_R}$  
on 
$\Tspace_R$,  
and a corresponding real unitary representation 
$\Pi_{\Kspace_R}$ 
on  
$\Kspace_R$, with the isomorphisms (\ref{isomorphisms}).  
 
\vs\ni 
Let $\AO$ denote the self-adjoint linear operator acting on $\Hspace$  
which generates the 1-parameter sub-representation of  
$\Pi_{\Hspace}$ corresponding to a 1-parameter subgroup $H<G$. 
Let $\alpha$ denote the self-adjoint linear operator acting on $\Kspace_R$ 
which generates the corresponding 1-parameter sub-representation of $\Pi_{\Kspace_R}$.  
If we are given  
$\Pi_{\Hspace}$, it is clear from (\ref{Kspacegroupaction2}) and (\ref{Kspacegroupaction3})
that we can determine $\Pi_{\Kspace_R}$,  
and so, given $\AO$, we can determine $\alpha$ in principle. We call this the `direct problem.' 
More interesting, and less obvious, is that given $\Pi_{\Kspace_R}$ and hence, implicitly, given  
$\alpha$, we can solve the `inverse problem' and determine $\AO$. 
In this way we attempt to determine, one 1-parameter subgroup at a time,  
the ray representation $\Pi_{\Hspace}$ from the real 
unitary representation $\Pi_{\Kspace_R}$, in effect  
performing the factorisation (\ref{isomorphisms}):
\be 
\Pi_{\Kspace_R}=\Pi_{\Hspace}\stackrel{W}{\otimes} \Pi^C_{\Hspace}\,.    
\label{factorisereps} 
\ee 
We consider two cases. In the first case, $(\AO-a{\hat I})\in\Tspace_R$ 
for some real constant $a$\,; in the second case, 
$\AO\in \Qspace_{WH}$.

\vs\ni 
Suppose firstly that we are given $(\AO-a{\hat I})\in\Tspace_R$ for some real $a$,  
and hence a corresponding function $A$ such that $(A-a)=\Wmap (\AO-a{\hat I}) \in\Kspace_R$.  
Let ${\tilde A}=A-a$.  
We look for $\alpha$ in the form of an integral operator \cite{moshinsky} on $\Kspace_R$\,: 
\be 
(\alpha B)(q_1,p_1) 
= 
\int \alpha_K(q_1,p_1,q_2,p_2) B(q_2,p_2)\,d\Gamma_2\,. 
\label{alphaaction1} 
\ee 
The local (Lie algebraic) condition corresponding to the global 
(group theoretic) condition (\ref{Kspacegroupaction2}) is  
\be 
\alpha B= A\star B -B\star A ={\tilde A}\star B-B\star{\tilde A}= i\{{\tilde A},B\}_M\,,
\label{localgenaction1} 
\ee 
from which it is easily checked that, as a consequence of (\ref{localgenaction1}),
\be
\alpha(B\star C)= (\alpha B)\star C + B\star (\alpha C)\,.
\label{alphaaction2}
\ee
This is the local condition corresponding to (\ref{AutVcondn2}).  
When (\ref{localgenaction1}) holds, we have from 
(\ref{moyal2}), for suitably smooth $B$, say $B\in\Jspace_R$,  
\bea 
\lefteqn{\int \alpha_K(q_1,p_1, q_2,p_2) B(q_2,p_2)\,d\Gamma_2 
} 
\mea 
&=&  
\frac{2i}{\pi^2} 
\int 
[\sin\{2[ 
p_1(q_2-q_3)+p_2(q_3-q_1) 
+p_3(q_1-q_2) 
]\} 
{\tilde A}(q_3,p_3)\,d\Gamma_3]\, 
\mea 
&\times & 
B(q_2,p_2) 
\,d\Gamma_2\,, 
\label{alphakernelcalc1} 
\eea 
and so  
\bea 
\lefteqn{\alpha_K (q_1,p_1, q_2,p_2)=\quad} 
\mea 
& & 
\frac{2i}{\pi^2} 
\int 
\sin\{2[ 
p_1(q_2-q_3)+p_2(q_3-q_1) 
+p_3(q_1-q_2) 
]\} 
{\tilde A}(q_3,p_3)\,d\Gamma_3\,. 
\mea 
\label{alphakernelcalc2} 
\eea 
Then (\ref{alphaaction1}) and (\ref{alphakernelcalc2}) define 
the action of $\alpha$ in terms of ${\tilde A}$ (and hence in terms of $A$ or 
$\AO$), thus solving the direct problem.  
Note that because ${\tilde A}$ is real, (\ref{alphakernelcalc2}) implies 
\be 
\overline{\alpha_K(q_1, p_1, q_2, p_2)} 
= 
\alpha_K(q_2, p_2, q_1, p_1)=-\alpha_K(q_1,p_1,q_2,p_2)\,, 
\label{alphaconstraints1} 
\ee 
as required by selfadjointness of $\alpha$, and the reality of $\Pi_{\Kspace_R}$.  
\vs\ni 
To solve the inverse problem, we must invert (\ref{alphakernelcalc2}).  
This will only be possible if $\alpha_K$ is further constrained, 
because the conditions 
(\ref{alphaconstraints1}) only guarantee that $\alpha$ generates an element of 
$O(\Kspace_R)$,
and we require that $\alpha$ generates an element of $\AutV<O(\Kspace_R)$.  
The further constraint is (\ref{alphaaction2}), 
but we wish to express it as a condition on $\alpha$ alone.
We change variables in 
(\ref{alphakernelcalc2}) and write it in the form 
\bea 
\lefteqn{\alpha_K((u'-u)/2, (v'+v)/2, (u'+u)/2, (v'-v)/2)} 
\mea 
&=&-\frac{2i}{\pi^2}\int \sin\{2v(q_3-u'/2)+2u(p_3-v'/2)\}{\tilde A}(q_3,p_3)\,d\Gamma_3 
\mea 
&=&-\frac{2i}{\pi^2}\int \sin\{2vq_3+2up_3\}{\tilde A}(q_3+u'/2,p_3+v'/2)\,d\Gamma_3 
\mea 
&=&-\frac{i}{2\pi^2}\int \sin\{vx+uy\}{\tilde A}((u'+x)/2,(v'+y)/2)\,dx\,dy 
\label{uvintro} 
\eea 
where  
\bea 
u=q_2-q_1\,,&\quad& v=p_1-p_2\,, 
\mea 
u'=q_2+q_1\,,&\quad& v'=p_1+p_2\,.  
\label{uvdef} 
\eea 
Now (\ref{uvintro}) takes the form  
\be 
R(u,v,u',v')=\int\sin(vx+uy) S(x,y,u',v')\,dx\,dy\,,  
\label{RSeqn} 
\ee 
where 
\bea 
R(u,v, u',v')&=& 
\alpha_K((u'-u)/2, (v'+v)/2, (u'+u)/2, (v'-v)/2) 
\mea 
S(x,y,u',v')&=&-\frac{i}{2\pi^2}{\tilde A}((u'+x)/2,(v'+y)/2)\,. 
\label{RSdef} 
\eea 
\vs\ni 
Set  
\be 
S^{(\pm)}(x,y,u',v')=\frac{1}{2}(S(x,y,u',v')\pm S(-x,-y,u',v'))\,, 
\label{Spmdef} 
\ee 
and note that (\ref{RSeqn}) can be written as 
\bea 
R(u,v,u',v')&=& \int\sin(vx+uy)S^{(-)}(x,y,u',v')\,dx\,dy 
\mea 
&=& -i\int e^{i(vx+uy)}S^{(-)}(x,y,u',v')\,dx\,dy\,. 
\label{RfourierS} 
\eea 
Inverting the double Fourier transform, we have 
\bea 
S^{(-)}(x,y,u',v') 
&=&\frac{i}{(2\pi)^2}\int e^{-i(vx+uy)} R(u,v,u',v')\,du\,dv 
\mea 
&=&\frac{1}{(2\pi)^2}\int \sin(vx+uy) R(u,v,u',v')\,du\,dv\,, 
\label{SReqn} 
\eea 
using  
$R(-u,-v,u',v')=-R(u,v,u',v')$, which follows from  
(\ref{alphaconstraints1}). 
Reintroducing $A$ from (\ref{RSdef}) and (\ref{Spmdef}), we have  
\bea 
\lefteqn{ 
A((u'+x)/2,(v'+y)/2) 
-A((u'-x)/2,(v'-y)/2)} 
\mea 
&=&2i\int \sin(vx+uy) 
R(u,v,u',v')\,du\,dv\,,  
\label{Aalphaeqn1} 
\eea 
and so  
\bea 
\lefteqn{ 
A(x,y) 
-A(0,0)} 
\mea 
&=&2i\int \sin(vx+uy)R(u,v,x,y) 
\,du\,dv\,. 
\label{Aalphaeqn2} 
\eea 
Then  
\bea 
\lefteqn{ 
A((u'+x)/2,(v'+y)/2) 
-A(0,0)=} 
\mea 
&&2i\int \sin(v(u'+x)/2+u(v'+y)/2)R(u,v,(u'+x)/2,(v'+y)/2) 
\,du\,dv\,,  
\mea 
\label{Aalphaeqn3} 
\eea 
with a similar expression for  
$A((u'-x)/2, (v'-y)/2)-A(0,0)$. Subtracting this second expression from the first, 
and equating to the RHS of (\ref{Aalphaeqn1}), we get  
\bea 
\lefteqn{ 
\int\sin(vx+uy) 
R(u,v,u',v')\,du\,dv=} 
\mea 
&&  
\int\sin(v(u'+x)/2+ u(v'+y)/2) 
R(u,v,(u'+x)/2,(v'+y)/2)\,du\,dv 
\mea 
&-&\int\sin(v(u'-x)/2+ u(v'-y)/2) 
R(u,v,(u'-x)/2, (v'-y)/2)\,du\,dv\,. 
\mea 
\label{Rcondn1} 
\eea 
It is this condition, with $R$ as in (\ref{RSdef}), that $\alpha_K$ must satisfy in addition to  
(\ref{alphaconstraints1}), if $\alpha$ is to generate an element of $\AutV<O(\Kspace_R)$. 
To see this, and to solve the inverse problem, suppose 
now that we are given $\alpha_K$ satsifying (\ref{alphaconstraints1}) 
and (\ref{Rcondn1}), with $R$ as in (\ref{RSdef}), and  
$u,v,u',v'$ as in (\ref{uvdef}).  
\vs\ni 
Set  
\be 
A(x,y)=a+2i\int\sin (vx+uy)R(u,v,x,y)\,du\,dv\,, 
\label{ARdef} 
\ee 
where $a$ is an arbitrary real constant, and check that $A$ is real, and that  
(\ref{Aalphaeqn1}) is satisfied. Then retrace the steps to recover 
(\ref{alphakernelcalc2}), showing that $A$ 
as given by (\ref{ARdef})  
generates the automorphism  
associated with $\alpha$.  
Note that $A$ is only defined by $\alpha$ up to the 
arbitrary real constant $a$, so the the corresponding  
unitary operator in $\Pi_{\Hspace}$ is only defined up to a constant phase, as expected.  
\vs\ni 
The treatment of this first case, with $\AO\in\Tspace_R$ and $A\in\Kspace_R$,  
might be extended to the case of a general selfadjoint 
$\AO$ and corresponding $\alpha$, with a suitable extension of  
the interpretation of the integral formulas above to accommodate distributions.  
We only consider further the second case mentioned above, 
when $\AO\in\Qspace_{WH}$. This can be treated more directly.  
 
\vs\ni 
Suppose then that  
$\AO$ is a hermitian polynomial in the canonical operators 
$\qO$, $\pO$ and ${\hat I}$, as introduced in Section 2. 
The corresponding $A$ is a real polynomial in $q$, $p$ and $1$ of the same degree, and  
according to (\ref{localgenaction1}) and (\ref{moyalB}), $\alpha$ is a polynomial in  
$q$, $p$, $\partial / \partial q$ and $\partial/\partial p$  
acting on suitably smooth $B\in\Kspace_R$ (say $B\in\Jspace_R$). This last polynomial 
is also of the same degree, except that it has no constant term. 
For example, corresponding to $A=q+a$, we have  
\be 
\alpha B = i\{q+a,B\}_{GM}=i\frac{\partial B}{\partial p}\,, 
\label{qcorresp} 
\ee
using (\ref{starprodWH}),  
so that $A=q+a$ corresponds to $\alpha =i\partial/\partial p$ for all values of the constant $a$. 

\vs\ni 
When restricted to act on an invariant subspace of $\Hspace$, 
the selfadjoint operators in $Q_{WH}$,  
\be 
\XO_1={\hat I}\,,\quad\XO_2=\qO\,,\quad\XO_3=\pO\,,
\quad \XO_4=\qO^2\,,\quad \XO_5=\frac{1}{2}(\qO\pO+\pO\qO)\,,\,\dots 
\label{Ldefn} 
\ee 
span an infinite-dimensional real Lie algebra ${\cal L}$.  
Choosing the coordinate representation 
$\Hspace\cong L_2({\mathbb C},dx)$ as in Section 2, we have the 
representation $\xi$ of ${\cal L}$ on $\Gspace<\Hspace$ with  
\bea 
\xi(\XO_1)=1\,,\quad \xi(\XO_2)=x\,, 
\quad\xi(\XO_3)=-i\frac{\partial}{\partial x}\,, 
\mea 
\xi(\XO_4)=x^2\,,\quad \xi(\XO_5)=-i(x\frac{\partial}{\partial x}+\frac{1}{2})\,,\dots 
\label{xirep} 
\eea 
The mapping $\Xi$,  
carrying selfadjoint operators $\AO$ in $Q_{WH}$ into 
corresponding selfadjoint operators $\alpha$ acting on $\Kspace_R$, 
defines a representation of ${\cal L}$ on $\Jspace_R$.  
This can be seen explicitly from (\ref{localgenaction1}), which gives for any $B\in\Jspace_R$,  
\be 
\Xi(\XO_i) 
\Xi(\XO_j) 
B=-\{\Wmap(\XO_i),\{\Wmap(\XO_j),B\}_{GM}\}_{GM}\,, 
\label{chimapaction} 
\ee 
so that  
\bea 
[\Xi(\XO_i),\Xi(\XO_j)]B&=&\{\{\Wmap(\XO_i),\Wmap(\XO_j)\}_{GM},B\}_{GM} 
\mea 
&=&\{\Wmap([\XO_i,\XO_j]),B\}_{GM} 
\mea 
&=& 
\Xi([\XO_i,\XO_j])B\,, 
\label{jacobi1} 
\eea 
using the antisymmetry property of the Groenewold-Moyal bracket, and the associated Jacobi
identity.  
Thus, when all the generators of  
the representation $\Pi_{\Kspace_R}$ of the group 
$G$ belong to $Q_{WH}$, they provide a representation on $\Jspace_R$ of the Lie algebra of $G$.  
Using (\ref{localgenaction1}), we find explicitly that  
\bea 
\Xi(\XO_1)=0\,,\quad\Xi(\XO_2)=i\frac{\partial}{\partial p}\,,
\quad\Xi(\XO_3)=-i\frac{\partial}{\partial q}\,,\quad \Xi(\XO_4)=2iq\frac{\partial}{\partial p}\,, 
\,\dots 
\label{Lrep1} 
\eea 
In Table 1 we list some corresponding $\AO$, $A=\Wmap(\AO)$ and 
$\alpha= \Xi(\AO)$ obtained using (\ref{localgenaction1}). 
Note that every $\alpha$ is formally hermitian and 
pure imaginary, as required by the unitarity and 
reality of $\Pi$. The extension of the operators $\AO$ and $\alpha$ from hermitian polynomials on  
$\Gspace$ and $\Jspace_R$ respectively, to selfadjoint 
operators on appropriate domains in $\Hspace$ and $\Kspace_R$ 
is straightforward.

\begin{table} 
\begin{tabular}{||l|l|l||} \hline 
&&\\ 
$\,\,\qquad\AO$ & $\quad A$ & $\qquad\qquad\qquad\qquad\quad \quad\alpha$ \\ 
\hline \hline 
&&\\ 
&& 
$i(P_q\frac{\partial }{\partial p}- 
P_p\frac{\partial }{\partial q})$\\ 
&&\\ 
$\Wmap^{-1}(P(q,p))$ & $P(q,p)$ &  
$
-\frac{i}{3!4}( 
P_{qqq}\frac{\partial ^3}{\partial p^3} 
-3P_{qqp} 
\frac{\partial ^3}{\partial q\partial p^2} 
+3P_{qpp} 
\frac{\partial ^3}{\partial q^2\partial p} $\\
&&\\ 
&&$ 
- 
P_{ppp} 
\frac{\partial ^3}{\partial q^3})
+\frac{i}{5!4^2}(P_{qqqqq}\frac{\partial ^5}{\partial p^5} -\dots $\\ 
&&\\ 
\hline 
&&\\ 
${\hat I}$ & $1$ & $0$\\ 
&&\\ 
$\qO$ & $q$ & $i\frac{\partial}{\partial p}$\\ 
&&\\ 
$\pO$ & $p$ & $-i\frac{\partial}{\partial q}$\\ 
&&\\ 
$\qO^2$ & $q^2$ & $2i q\frac{\partial}{\partial p}$\\ 
&&\\ 
$\pO^2$ & $p^2$ & $-2i p\frac{\partial}{\partial q}$\\ 
&&\\ 
$\frac{1}{2}(\qO\pO + \pO\qO)$ & $qp$ & $ip\frac{\partial}{\partial p}-iq  
\frac{\partial}{\partial q} 
$  
\\ 
&&\\ 
$\qO^3$ & $q^3$ & $3iq^2 
\frac{\partial}{\partial p} 
- \frac{1}{4}i  
{\frac{\partial ^3}{\partial p^3}}$\\ 
&&\\ 
$\pO^3$ & $p^3$ & $-3ip^2 
\frac{\partial}{\partial q} 
+ \frac{1}{4}i  
{\frac{\partial ^3}{\partial q^3}}$\\ 
&&\\ 
$\qO\pO\qO$ & $q^2p$ & $ 
2i qp 
\frac{\partial}{\partial p} 
-iq^2 
\frac{\partial}{\partial q} 
+\frac{1}{8}i 
{\frac{\partial ^3}{\partial q\partial p^2}}$\\ 
&&\\ 
$\pO\qO\pO$ & $qp^2$ & $ip^2 
\frac{\partial}{\partial p} 
-2i qp 
\frac{\partial}{\partial q} 
-\frac{1}{8}i 
{\frac{\partial ^3}{\partial q^2\partial p}}$\\ 
&&\\ 
&&\\ 
$V(\qO)$ & $V(q)$ & $ iV^{(1)}(q)\frac{\partial}{\partial p} 
-\frac{i}{3! 4}V^{(3)}(q)\frac{\partial^3}{\partial p^3} $\\
&&\\
&&$+\frac{i}{5!4^2}V^{(5)}(q)\frac{\partial^5}{\partial p^5}
-\dots$\\ 
&&\\ 
\hline 
\end{tabular} 
\caption{Corrresponding $\AO$, $A$ and $\alpha$. 
Subscripts on $P$ indicate 
partial derivatives, and $V^{(n)}$ denotes 
the $n$-th derivative of $V$.}
\end{table}

\vs\ni 
Note that $\Xi$ defines a Lie algebra homomorphism but not an algebra homomorphism:
it does not define a representation of the enveloping algebra of the Heisenberg-Weyl Lie algebra.
For example, as can be seen from the Table, 
$
\Xi(\XO_2)\Xi(\XO_3)
+
\Xi(\XO_3)\Xi(\XO_2)
\neq
\Xi(\XO_2\XO_3 + \XO_3\XO_2)$.

\vs\ni
Corresponding to (\ref{isomorphism}), the representation 
$\Xi$ of ${\cal L}$ is isomorphic to the tensor product
of the representation $\xi$ on $L_2({\mathbb C},dx)$ as in (\ref{xirep}) 
and its contragredient $\xi^C$ on $L_2({\mathbb C},dy)$, so that 
on $\Jspace_R$,
\be
\Xi=Z\,(\xi\otimes\xi^c)\,Z^{\dagger}\,, 
\label{isomorphism} 
\ee 
where $Z$ is the unitary transformation (\ref{Zdef}),  
and  
\bea 
\xi^C(\XO_1)=-1\,,\quad\xi^C(\XO_2)=-y\,,\quad 
\xi^C(\XO_3)=-i\frac{\partial}{\partial y}\,,\quad  
\mea 
\xi^C(\XO_4)=-y^2\,,\quad\xi^C(\XO_5)=-i(y\frac{\partial}{\partial y}+\frac{1}{2})\,,\,\dots\,. 
\label{xiCrep} 
\eea 
The rule in going from (\ref{xirep}) to (\ref{xiCrep}) is that  
each real expression attracts a minus sign, whereas each pure imaginary expression does not.  
Straightforward calculation shows that 
\bea 
Z\,x\,Z^{\dagger}=q+\frac{1}{2}i\frac{\partial}{\partial p}\,,\quad 
Z\,y\,Z^{\dagger}=q-\frac{1}{2}i\frac{\partial}{\partial p}\,,\quad 
\mea 
Z\,i\frac{\partial}{\partial x}\,Z^{\dagger}=-p+\frac{1}{2}i\frac{\partial}{\partial q}\,,\quad 
Z\,i\frac{\partial}{\partial y}\,Z^{\dagger}=p+\frac{1}{2}i\frac{\partial}{\partial q}\,, 
\label{Zonpolys1} 
\eea 
with inverses 
\bea 
Z^{\dagger}\,q\,Z=\frac{1}{2}(x+y)\,,\quad 
Z^{\dagger}\,i\frac{\partial}{\partial p}\,Z=x-y\,,\quad 
\mea 
Z^{\dagger}\,p\,Z=\frac{1}{2}( 
-i\frac{\partial}{\partial x} 
+i\frac{\partial}{\partial y})\,,\quad 
Z^{\dagger}\,i\frac{\partial}{\partial q}\,Z= 
i\frac{\partial}{\partial x} 
+i\frac{\partial}{\partial y}\,, 
\label{Zonpolys2} 
\eea 
from which one can easily deduce that, corresponding to 
the monomial $A=q^mp^n$ and its image $\AO$ under $\Winv$ as in (\ref{berezin2}), we have  
\bea 
\lefteqn{\Xi(\Winv(q^mp^n))} 
\mea 
&=& \frac{1}{2^m}\sum_{r=0}^m\,C^m_r\,[ 
(q+\frac{1}{2}i\frac{\partial}{\partial p}) 
^{m-r} 
(p-\frac{1}{2}i\frac{\partial}{\partial q}) 
^n 
(q+\frac{1}{2}i\frac{\partial}{\partial p}) 
^r  
\mea 
&-& (q-\frac{1}{2}i\frac{\partial}{\partial p}) ^{m-r} 
(p+\frac{1}{2}i\frac{\partial}{\partial q})
^n
(q-\frac{1}{2}i\frac{\partial}{\partial p})
^r]\,.
\label{genmonomial2}
\eea
Related formulas have been presented recently  by Hakioglu and Dragt
\cite{dragt}. 
\vs\ni 
The extra constraint corresponding to 
(\ref{Rcondn1}), which $\alpha$ must satisfy in order to generate an element of $\AutV$, 
is simply this: only those polynomials in 
$q$, $p$, $\partial / \partial q$ and $\partial/\partial p$  
which are real linear combinations of those in (\ref{genmonomial2}),  
represent possible $\alpha$. For example, $\alpha=iq^2\partial/\partial p$ is not allowed; it  
generates 
an element of $O(\Kspace_R)$ that is not in $\AutV$.  
A more straightforward test of a candidate $\alpha$  
is to evaluate $(Z^{\dagger}\,\alpha\,Z)$ using (\ref{Zonpolys2}).  
The resulting operator on $L_2({\mathbb C},dx)\otimes L_2({\mathbb C},dy)$ must have the form  
$\AO(x,-i\partial/\partial x)-\overline{\AO(y,-i\partial/\partial y)}$ for some  
hermitian polynomial operator $\AO(x,-i\partial/\partial x)$ on $L_2({\mathbb C},dx)$.  
 
\vs\ni 
The direct and inverse problems in this case 
are solved simply by consulting Table 1, extended if necessary to higher degrees 
using (\ref{localgenaction1}) or (\ref{genmonomial2}). 
That is to say, given $\AO$ (or $A$), read off 
$\alpha$; given $\alpha$, read off $\AO$ (or $A$) up to the addition 
of an arbitrary real constant multiple of ${\hat I}$ (or $1$).  
Alternatively, to solve the inverse problem, proceed as 
in the preceding paragraph to identify $\AO$ (to within a constant multiple of the identity 
operator).

\section{Examples}
\ni
It is informative in the first two examples to use  
dimensional variables, introducing factors of $\hbar$ in the appropriate places.
\vs\ni

\vs\ni 
1. \quad{\it The Heisenberg-Weyl group}  
\vs\ni 
Elements of this 2-parameter, real, Abelian Lie group are labelled $g(a_1,a_2)$,  
where $a_1$ and $a_2$ take all real values, and the product rule is  
\be 
g(a_1,a_2)g(b_1,b_2)=g(a_1+b_1,a_2+b_2)\,,  
\label{WHgroupaction} 
\ee 
The real, true, unitary representation on $\Kspace_R$ in this case has the form 
\be 
(\Pi_{\Kspace_R}(a_1,a_2)F)(q,p) = F(q+a_1,p-a_2)\,,  
\label{WHaction1} 
\ee 
with the associated generators  
\be 
\alpha_1=-i\frac{\partial}{\partial q}\,,\quad 
\alpha_2=i\frac{\partial}{\partial p}\,,  
\label{WHphasegens} 
\ee 
satisfying  
on $\Jspace_R$  
the commutation relation  
\be 
[\alpha_1,\alpha_2]=0\,. 
\label{WHcommutator1} 
\ee 
Note that Planck's constant does not appear in $\Pi_{\Kspace_R}$.
\vs\ni 
Turning to the `factorisation' (\ref{factorisereps}), 
we have from (\ref{WHaction1}) and (\ref{Zinvdef}) that
\bea
(Z^{\dagger}\Pi_{\Kspace_R}(a_1,a_2)F)(x,y)=
\frac{1}{2\pi\hbar}
\int
F(\frac{x+y}{2}+a_1,p-a_2)e^{ip(x-y)/\hbar}\,dp&&
\mea
=
\frac{1}{2\pi\hbar}
e^{ia_2(x-y)/\hbar}
\int
F(\frac{x+y}{2}+a_1,p)e^{ip(x-y)/\hbar}\,dp\,.&&
\label{heiseval1}
\eea
Setting
\be
(Z^{\dagger}F)(x,y)=f(x,y)\,,
\label{heiseval2}
\ee
so that, from (\ref{Zinvdef}), 
\be
f(x,y)=
\frac{1}{2\pi\hbar}
\int
F(\frac{x+y}{2},p)e^{ip(x-y)/\hbar}\,dp\,,   
\label{heiseval3}
\ee
we have from (\ref{heiseval1}) that
\be
((\Pi_{L_2({\mathbb C},dx)}
\otimes
\Pi_{L_2({\mathbb C},dy)}^C)f)(x,y)=e^{ia_2(x-y)/\hbar}f(x+a_1,y+a_1)\,,
\label{heiseval4}
\ee
From this we see that a possible factorisation is obtained by taking 
the action of $\Pi_{L_2({\mathbb C},dx)}$ and 
$\Pi_{L_2({\mathbb C},dy)}^C$ on $u\in L_2({\mathbb C},dx)$ and $v\in L_2({\mathbb C},dy)$, 
respectively, to be 
\bea 
(\Pi_{L_2({\mathbb C},dx)}(a_1,a_2)u)(x)&=&e^{i\omega(a_1,a_2)} e^{ia_2x/\hbar}u(x+a_1)\,, 
\mea 
\mea 
(\Pi^C_{L_2({\mathbb C},dy)}(a_1,a_2)v)(y)&=&e^{-i\omega(a_1,a_2)} e^{-ia_2y/\hbar}v(y+a_1)\,. 
\label{WHaction2} 
\eea 
where $\omega$ is real-valued. It is then readily checked that 
$\Pi_{L_2({\mathbb C},dx)}$ is a projective representation of the Abelian group (\ref{WHgroupaction}), 
whatever the form of  
$\omega$. Different choices for $\omega$ 
correspond to different choices, from the same cohomology class,  
of the cocycle associated with projective representations of the group, and do not differ in a 
significant way. We may say that,  
up to the phase $\omega$, we have recovered in 
(\ref{WHaction2}) the usual projective unitary representation on $\Hspace$,
realised as $L_2({\mathbb C},dx)$.  

\vs\ni
We can also consider this example from the Lie algebraic 
viewpoint. Using (\ref{Zonpolys2}), now with appropriate factors of $\hbar$ inserted,  
we have at once from 
(\ref{WHphasegens}) that  
\be 
Z^{\dagger}\,\alpha_1\,Z=\,\quad -i\frac{\partial}
{\partial x}-i\frac{\partial}{\partial y}\,,\quad 
Z^{\dagger}\,\alpha_2\,Z=\frac{x}{\hbar}-\frac{y}{\hbar}\,, 
\label{WHhilbertgens1} 
\ee 
from which we have  
\be 
\AO_1=\pO=-i\hbar\frac{\partial}{\partial x}+p_0\,,\,\,\,
\AO_2=\qO=x+q_0\,,  
\label{WHhilbertgens2} 
\ee 
where $q_0$ and $p_0$ are arbitrary constants.  
Then 
\be 
[\qO,\pO]=i\hbar
\label{canonicalops1} 
\ee 
on $\Gspace$, and $\AO_1$, $\AO_2$ 
are equivalent to the usual canonical operators there. 
Note that $\hbar$ appears on the RHS of (\ref{canonicalops1}) 
as the parameter associated with a central extension of the Lie algebra  
in going from (\ref{WHcommutator1}) to (\ref{canonicalops1}). 
There is no $\hbar$ in $\Pi_{\Kspace_R}$, but there 
is in $\Pi_{\Hspace}$. Evidently its appearance comes 
from the unitary transformation $Z$, or equivalently, from the 
Weyl-Wigner transform. Note also that from this point of view, 
$\hbar$ is an arbitrary parameter; the factorisation  
(\ref{factorisereps}) of $\Pi_{\Kspace_R}$ works for any value of $\hbar$ in $Z$ in this case, 
and the one to be chosen ultimately is a matter for physics to decide.  

\vs\ni
The Heisenberg-Weyl algebra can be generalised in an obvious way, showing that
Lie algebras with polynomial elements of arbitrarily high degree
can arise in the phase-space formalism.  Consider the $N+1$-dimensional real Lie algebra with 
selfadjoint representation generated by 

\be
\alpha_1=
-i\frac{\partial}{\partial q}\,,
\label{genheisdef1}
\ee
together with 

\bea
\beta_1=
i\frac{\partial}{\partial p}\,,\quad
\beta_2=
iq\frac{\partial}{\partial p}\,,\quad
\beta_3=
\frac{1}{2} iq^2\frac{\partial}{\partial p}
-\frac{1}{24}i\frac{\partial ^3}{\partial p^3}\,,\,\dots\beta_N\,,\quad{\rm that}\,\,\,{\rm is}&&
\mea
\mea
\beta_n=\frac{2}{n!}\sum_{m=1}^n
\left(\frac{i}{2}\right)^{n-m}
C^n_mq^m\frac{\partial^{n-m}}{\partial p^{n-m}}\,,\,n=1,\,2,\,\dots\,N \,,&&
\label{genheisdef2}
\eea
where $C_m^n$ is the binomial coefficient as in (\ref{berezin1}), and 
the sum is restricted to odd values of $n-m$. 
In generalisation of (\ref{WHhilbertgens2}), we find that
the corresponding operators on $\Hspace$ are 
\be
\AO_1=
-i\hbar\frac{\partial}{\partial x}\,,\quad \BO_n=\frac{x^n}{n!}\,,\,n=1,\,2,\,\dots\,N\,,
\label{genheisdef3}
\ee
up to the addition of arbitrary real constants.   Once again, the Lie algebra generated by
$\AO_1$ and the $\BO_n$ is a central extension of the Lie algebra generated by $\alpha_1$ and
the $\beta_n$,
and Planck's constant appears as the extension parameter.

\vs\ni
2. \quad{\it The Galei group.} 

\vs
\ni
For a system with one degree of freedom, this 
group is a 3-parameter real Lie group \cite{leblond} 
with elements $g(a_1,a_2,a_3)$, 
where $a_1$, $a_2$ and $a_3$ take all real values, and the
product rule is 
\be
g(a_1,a_2,a_3)g(b_1,b_2,b_3)= g(a_1+b_1, a_2+b_2, a_3+b_3+b_2a_1)\,.
\label{GGproductrule}
\ee
Consider the true, real, unitary representation on $\Kspace_R$ defined by 
\be
(\Pi_{\Kspace_R}(a_1,a_2,a_3)F)(q,p)=F(q-\frac{a_1}{m}p-a_2a_1-a_3,p+ma_2)\,,
\label{galileiaction1}
\ee
with associated generators
\be
\alpha_1=-i\frac{p}{m}\frac{\partial}{\partial q}\,,\quad 
\alpha_2= im\frac{\partial}{\partial p}\,,\quad
\alpha_3=-i\frac{\partial}{\partial q}
\label{galKgens}
\ee
satisfying the commutation relations
\be
[\alpha_1,\alpha_2]=-i\alpha_3\,,\quad 
[\alpha_2,\alpha_3]=0\,,\quad 
[\alpha_1,\alpha_3]=0\,.
\label{galCRs1}
\ee
Note in this case that $m$ appears as a parameter in the  action
(\ref{galileiaction1}) 
of the group representation,
though not in the commutation relations (\ref{galCRs1}).
\vs\ni
It is easiest to perform a factorisation 
in this case after realising $\Hspace$ and its dual
as $L_2({\mathbb C},dr)$ and $L_2({\mathbb C},ds)$, respectively,
where $r$ and $s$ are `momentum' variables. In place of  (\ref{Zdef}) and
(\ref{Zinvdef}), we have
\bea
F(q,p)&=&
\int
{\tilde f}(p-\frac{r}{2}, p+\frac{r}{2})e^{-irq/\hbar}\,dr
=({\tilde Z}{\tilde f})(q,p)\,,
\mea
{\tilde f}(r,s)&=&
\frac{1}{2\pi}
\int
F(q,\frac{r+s}{2})e^{-iq(r-s)/\hbar}\,dq
=({\tilde Z}^{\dagger}F)(r,s)\,. 
\label{momentumtransforms}
\eea
Considering (\ref{galileiaction1}), we then have 
\bea
({\tilde Z}^{\dagger}\Pi_{\Kspace_R}F)(r,s)=\qquad\qquad\qquad
\mea\mea
\frac{1}{2\pi}
\int
F(q-\frac{a_1}{m}\frac{r+s}{2}-a_2a_1-a_3,\frac{r+s}{2}+ma_2)e^{-iq(r-s)/\hbar}\,dq\,,
\label{PiF1}
\eea
so that 
\bea
((\Pi_{L_2({\mathbb C},dr)}\otimes\Pi^C_{L_2({\mathbb C},ds)}){\tilde f})(r,s)=
({\tilde Z}^{\dagger}\Pi_{\Kspace_R}F)(r,s)=\qquad\qquad\qquad\qquad
\mea\mea
\frac{1}{2\pi}
\int
{\tilde f}
(\frac{r+s}{2}+ma_2-\frac{r'}{2},
\frac{r+s}{2}+ma_2+\frac{r'}{2})\qquad\qquad\qquad
\mea
\mea
e^{-ir'(q+\frac{-a_1}{m}\frac{r+s}{2}-a_2a_1-a_3)/
\hbar}e^{-iq(r-s)/\hbar}\,dq\,dr'\,,\qquad\qquad
\mea\mea
={\tilde f}(r+ma_2,s+ma_2)e^{-i\frac{a_1}{m}(r^2-s^2)}
e^{-i(a_2a_1+a_3)(r-s)}\,.\qquad\qquad\qquad
\label{PiF3}
\eea
We see that a possible factorisation has 
\bea
(\Pi_{L_2({\mathbb C},dr)}(a_1,a_2,a_3)u)(r)=
e^{i\omega(a_1,a_2,a_3)}e^{-i\frac{a_1}{m}r^2}e^{-i(a_2a_1+a_3)r}u(r+ma_2)
\mea
(\Pi_{L_2({\mathbb C},ds)^C}(a_1,a_2,a_3)v)(s)=
e^{-i\omega(a_1,a_2,a_3)}e^{i\frac{a_1}{m}s^2}e^{i(a_2a_1+a_3)s}v(s+ma_2)\,,
\label{galfactorise}
\eea
which, up to the arbitrary phase $\omega$, is the familiar action of the unitary ray
representation of the Galilei group in the momentum space realisation of
Hilbert space, and of its contragredient representation.

\vs\ni
From the Lie algebraic viewpoint, we find from (\ref{Zonpolys2})
\bea
Z^{\dagger}\,\alpha_1\,Z=
-\frac{\hbar}{2m} \frac{\partial ^2}{\partial x^2}
+\frac{\hbar}{2m} \frac{\partial ^2}{\partial y^2}\,,&\quad&
Z^{\dagger}\,\alpha_2\,Z=\frac{m}{\hbar} x-\frac{m}{\hbar}y\,,
\mea
Z^{\dagger}\,\alpha_3\,Z&=&
-i\frac{\partial}{\partial x}
-i\frac{\partial}{\partial y}\,,
\label{galHgens1}
\eea
from which we deduce that, in the coordinate reopresentation now, 
\bea
\AO_1=\HO=
-\frac{\hbar ^2}{2m} \frac{\partial ^2}{\partial x^2}+ e_0\,,&\quad&
\AO_2={\hat K}=m(x+q_0)\,,
\mea
\AO_3=\pO=
-i\hbar\frac{\partial}{\partial x}+p_0\,,
\label{galHgens2}
\eea
where $e_0$, $q_0$ and $p_0$ are arbitrary constants. 
Then $\HO$, ${\hat K}$ and 
$\pO$ are equivalent to the usual Hamiltonian,
boost and momentum operators for the free particle 
in one dimension, and satisfy on $\Gspace$ the familiar relations
\be
[\HO,{\hat K}]=-i\hbar \pO\,,\quad
[\HO,\pO]=0\,,\quad
[{\hat K},\pO]=\hbar m\,.
\label{galCRs2}
\ee
Comparing with (\ref{galCRs1}), we see the appearance 
of $m$ in (\ref{galCRs2}), associated with a central extension of the Lie algebra
of the Galilei group. Although $\Pi_{\Kspace_R}$ 
is a true representation of the group, associated with the commutation
relations (\ref{galCRs1}) 
in which no $m$ appears, nevertheless $m$ is a parameter 
in $\Pi_{\Kspace_R}$, enabling the factorisation 
(\ref{galfactorise}) to take place, and the $m$ to appear in (\ref{galCRs2}).

\vs\ni
3.\quad{\it Two 1-parameter groups}
\vs\ni
Consider the 1-parameter transformations group acting on 
$\Gamma$ with generator $\alpha$, whose kernel as in (\ref{alphaaction1}) is given by  
\bea
\lefteqn{\alpha_K(q_1,p_1,q_2,p_2)=}
\mea
&&
i\sin[(1+\epsilon)(p_1q_2-p_2q_1)-(1-\epsilon)(q_1p_1-q_2p_2)]
e^{-(q_1-q_2)^2/\tau -(p_1-p_2)^2/\sigma}
\mea
&\Rightarrow &R(u,v,u',v')=i\sin(uv'+\epsilon u'v)e^{-u^2/\tau-v^2/\sigma}\,,
\label{alphaKex}
\eea
where $\tau$ and $\sigma$ are positive constants, and $\epsilon=\pm 1$. 
Then the constraints (\ref{alphaconstraints1}) are satisfied, 
so that the corresponding operator $\alpha$ generates 
an element of $O(\Kspace_R)$.
Now we have 
\bea
\lefteqn{2i\int \sin(vx+uy)R(u,v,u',v')\,du\,dv = }
\mea
&&2\pi\sqrt{\tau\sigma}\left(
e^{-
((y+v')/2)^2\tau} e^{-((x+\epsilon u')/2)^2\sigma}
- 
e^{-
((y-v')/2)^2\tau} e^{-((x-\epsilon u')/2)^2\sigma}\right)\,,
\mea
\label{exRint}
\eea
and (\ref{Rcondn1}) is seen to be satisfied if $\epsilon=1$, 
but not satisfied if $\epsilon=-1$.
If $\epsilon=1$, (\ref{ARdef}) gives
\be
A(q,p)=a+2\pi\sqrt{\sigma\tau} e^{-\sigma q^2 -\tau p^2}\,,
\label{exAsoln}
\ee
with $a$ arbitrary. If $\epsilon=-1$, the element of $O(\Kspace_R)$ 
generated by $\alpha$ is not an element of $\AutV$, and no $A$ exists.

\vs\ni 
4. \quad{\it The Lie algebra $sp(2,R)$: Case A.} 

\vs\ni 
We consider the representation on $\Kspace_R$ 
with
\bea
\alpha_1= \frac{i}{2}(p\frac{\partial}{\partial p}-q\frac{\partial}{\partial q})\,, 
\quad 
\alpha_2= \frac{i}{2}(q\frac{\partial}{\partial p}+p\frac{\partial}{\partial q})\,, 
\mea
\alpha_3= \frac{i}{2}(q\frac{\partial}{\partial p}-p\frac{\partial}{\partial q})\,, 
\label{sp2gens1}
\eea
satisfying
\be
[\alpha_1,\alpha_2]=-i\alpha_3\,,\quad [\alpha_2,\alpha_3]
=i\alpha_1\,,\quad [\alpha_3,\alpha_1]=i\alpha_2\,.
\label{sp2CRs1}
\ee
Performing the factorisation as in the previous examples, we get in this case
(up to the addition of constant terms, which can be removed by redefinitions)
\bea
\AO_1 =-\frac{i}{2}(x\frac{\partial}{\partial x}+1)\,,&\quad&
\AO_2=\frac{1}{4}(x^2+\frac{\partial ^2}{\partial x^2})\,,
\mea
\AO_3=\frac{1}{4}(x^2-\frac{\partial ^2}{\partial x^2})\,,
\label{sp2gens2}
\eea
satisfying relations corresponding to (\ref{sp2CRs1}). 
This is the representation associated with the simple harmonic oscillator.
Each $\AO_i$ is quadratic in the canonical operators on 
$\Gspace$, and the quadratic Casimir operator for the Lie algebra
has the value
\be
-\AO_1^2-\AO_2^2+\AO_3^2=-\frac{3}{16}\,. 
\label{sp2casimirA}
\ee
No non-trivial central extensions are involved in this case, 
which has been described elsewhere in the phase space context, from 
a slightly different point of view 
\cite{dragt}. 
\vs\ni
It is interesting in this example to compare the reductions to
irreducible components of the Weyl-Wigner product $\Pi_{\Kspace_R}$,  and
the usual tensor product
$\Pi_{\Hspace}\otimes\Pi_{\Hspace}^C$, which can be found explicitly in both representations.  
In $\Pi_{\Kspace_R}$ we look for the common eigenfunctions
of  
\bea
\Lambda^2&=& -\alpha_1^2-\alpha_2^2+\alpha_3^2+\frac{1}{4}
\mea
&=& \frac{1}{4}
\left(p^2\frac{\partial^2}{\partial p^2}
+3p\frac{\partial}{\partial p}
+q^2\frac{\partial^2}{\partial q^2}
+3q\frac{\partial}{\partial q}
+2pq\frac{\partial^2}{\partial q\partial p}+1\right)
\mea
&=&
(\frac{1}{2}(r\frac{\partial}{\partial r}+1))^2
\,,
\label{sp2casimirK}
\eea
and 
\be
\alpha_3= \frac{i}{2}(q\frac{\partial}{\partial p}-p\frac{\partial}{\partial q}) 
 =\frac{1}{2}i\frac{\partial}{\partial \theta}\,,
\label{alpha3polar}
\ee
where in (\ref{alpha3polar}) and the last line of (\ref{sp2casimirK}) we have introduced polar
variables in the phase plane: 
\bea
q=r\cos(\theta)\,,&& p=r\sin(\theta)\,.
\mea
0\leq r\leq \infty\,,&& 0\leq \theta <2\pi\,.
\label{polars}
\eea
The common (generalised, unnormalised) eigenfunctions of $\Lambda$ and $\alpha_3$ are   
then seen to be 
\bea
\Phi_{\lambda,m}(r,\theta)&=&\frac{e^{-i\lambda\ln r}}{r}\,e^{-im\theta}
\mea
-\infty<\lambda <\infty\,,&&m=0,\,\pm1,\,\pm2,\,\dots\,.
\label{Keigenfunctions}
\eea
In fact there are two irreducible
representations here for each value of $\lambda$, one with all even integer values of
$m$, and one with all odd integer values.

\vs\ni
In $\Pi_{\Hspace}\otimes \Pi_{\Hspace}^C$, with $\AO_i(x, -i\partial /\partial x)
=\AO_i$ as in (\ref{sp2gens2}),
we seek the common eigenfunctions $\Psi_{\lambda',m'}(x,y)$ of 
\bea
-\left(\AO_1(x,-i\partial /\partial x)-
\overline{\AO_1(y, -i\partial /\partial y)}\right)^2 &&
\mea
-\left(\AO_2(x,-i\partial /\partial x)-
\overline{\AO_2(y, -i\partial /\partial y)}\right)^2&& 
\mea
+\left(\AO_3(x,-i\partial /\partial x)-
\overline{\AO_3(y, -i\partial /\partial y)}\right)^2 
+\frac{1}{4}&&
\mea
=
-\frac{1}{4}[i(ab-a^{\dagger}b^{\dagger})]^2=-\Lambda'^2\,,
\quad{\rm say}\,,&&
\mea
\label{sp2casimirHH}
\eea
and
\be
J_3=
\AO_3(x)-\AO_3(y)
=\frac{1}{2}(a^{\dagger}a-b^{\dagger}b)\,,
\label{J3boson}
\ee
where we have introduced the boson operators
\bea
a=\frac{1}{\sqrt{2}}(x+\frac{\partial}{\partial x}),
&&
b=\frac{1}{\sqrt{2}}(y+\frac{\partial}{\partial y}),
\mea
\mea
a^{\dagger}=\frac{1}{\sqrt{2}}(x-\frac{\partial}{\partial x}),
&&
b^{\dagger}=\frac{1}{\sqrt{2}}(y-\frac{\partial}{\partial y})\,.
\label{bosondefs}
\eea
When $m'$ is nonnegative, these eigenfunctions have
the (unnormalised) form 
\be
\Psi_{\lambda ',m'}(x,y)=a^{\dagger m'} W_{\frac{\lambda '}{2},\frac{m'}{2}}
(2a^{\dagger}b^{\dagger})\varphi_0(x,y)\,,
\label{boseefns}
\ee
where $\varphi_0(x,y)=\exp [-(x^2+y^2)/2]$ is the `vacuum vector,' 
annihilated by $a$ and $b$, and $W_{\mu\nu}$
denotes a Whittaker function \cite{Abram}. 
When $m'$ is negative, the prefactor $a^{\dagger m'}$ on the RHS must be replaced by $b^{\dagger
-m'}$.  
Again there are two irreducible representations here for each value
of $\lambda '$, one with all even integer values of $m'$, and one with all odd integer values.   
The basis functions $\Phi_{\lambda,m}(q,p)$ and 
$\Psi_{\lambda,m}(x,y)$ must of course be related as in
(\ref{Zdef}) and (\ref{Zinvdef}), 
but it is by no means obvious that this is so.

\vs\ni
5. \quad{\it The Lie algebra $sp(2,R)$: Case B.} 

\vs\ni
As another  example where generators  of higher 
degree than quadratic in the underlying variables occur,
we consider the selfadjoint representation of $sp(2,R)$ on $\Kspace_R$ 
with
\bea
\alpha_1=\frac{1}{2}i
\frac{\partial }{\partial p}
-\frac{1}{2}ip^2
\frac{\partial }{\partial p}
+i(qp-\frac{1}{2}a)
\frac{\partial }{\partial q}
+\frac{1}{8}i
\frac{\partial^3 }{\partial q^2\partial p}
\,,\,\,\, 
\mea
\mea
\alpha_2 = -i
(q\frac{\partial }{\partial q}
-p\frac{\partial }{\partial p})
\,,
\mea
\mea
\alpha_3=-\frac{1}{2}i
\frac{\partial }{\partial p}
-\frac{1}{2}ip^2
\frac{\partial }{\partial p}
+i(qp-\frac{1}{2}a)
\frac{\partial }{\partial q}
+\frac{1}{8}i
\frac{\partial^3 }{\partial q^2\partial p}\,,
\label{sp2gens3} 
\eea 
again satisfying (\ref{sp2CRs1}), with $a$ an arbitrary real parameter. 
Performing the factorisation, 
we get in this case (again after redefinitions, where necessary)
\bea
\AO_1 =
\frac{1}{2}(x+x\frac{\partial ^2}{\partial x^2}+(1-ia)\frac{\partial}{\partial x})
\,,&\quad&
\AO_2=-i(x\frac{\partial }{\partial x}+\frac{1}{2}(1-ia))\,,
\mea
\mea
\AO_3 =
-\frac{1}{2}(x-x\frac{\partial ^2}{\partial x^2}-(1-ia)\frac{\partial}{\partial x})
\label{sp2gens4}
\eea
satisfying relations corresponding to (\ref{sp2CRs1}). 
The quadratic Casimir in this case has the value
\be
-\AO_1^2-\AO_2^2+\AO_3^2=-\frac{1}{4}(a^2+1)\,, 
\label{sp2casimirB}
\ee
showing that, whatever the real value of $a$, this selfadjoint representation on $\Hspace$
is inequivalent to the one 
associated with (\ref{sp2gens2}) and (\ref{sp2casimirA}). 
Again, no non-trivial central extensions are involved in this case.

\vs\ni
5. \quad{\it Time reversal.} 
\vs\ni
The group has two elements $g$ and $e$ (identity) with $g^2=e$.
The real, true,
unitary representation $\Pi_{\Kspace_R}$ acts as 
\be
(\Pi_{\Kspace_R}(g)F)(q,p)=F(q,-p)\,,\quad 
(\Pi_{\Kspace_R}(e)F)(q,p)=F(q,p)\,, 
\label{timereversal1}
\ee
for every $F\in\Kspace_R$.    
We have from (\ref{Zinvdef}),
\bea
(Z^{\dagger}\Pi_{\Kspace_R}(g)F)(x,y)&=&
\frac{1}{2\pi}
\int
F(\frac{x+y}{2},-p)e^{ip(x-y)}\,dp\,,
\mea
&=&
\frac{1}{2\pi}
\int
F(\frac{x+y}{2},p)e^{-ip(x-y)}\,dp\,,
\label{timereversal6}
\eea
because $F$ is real.  
Defining $f(x,y)=(Z^{\dagger}F)(x,y)$ as in (\ref{Zinvdef}), we have 
\be
\overline{f(x,y)}=
\frac{1}{2\pi}
\int
F(\frac{x+y}{2},p)e^{-ip(x-y)}\,dp\,,
\label{timereversal2}
\ee
and so 
\be
((\Pi_{L_2({\mathbb C},dx)}(g)
\otimes
\Pi_{L_2({\mathbb C},dy)}(g)^C)f)(x,y)=
(Z^{\dagger}\Pi_{\Kspace_R}(g)F)(x,y)=
\overline{f(x,y)}\,.
\label{timereversal3}
\ee
Now it can be seen that a possible factorisation has 
\bea
((\Pi_{L_2({\mathbb C},dx)}(g)u)(x)&=&e^{i\omega}{\overline u}(x)\,,
\mea
((\Pi_{L_2({\mathbb C},dy)}^C(g)u)(x)&=&e^{-i\omega}{\overline v}(y)\,,
\label{timereversal4}
\eea
with $\omega$ any real number, so that
\be
\Pi_{\Hspace}(g)=\Pi_{\Hspace}^C(g)=e^{i\omega}\CO\,,
\label{timereversal5}
\ee
where $\CO$ is the antiunitary operator of (\ref{Caction}) and (\ref{genPaction1}). 

\vs
\centerline{*\qquad\qquad*\qquad\qquad*}
\vs\ni
AJB thanks GC for his generous hospitality, and 
R.C. King for useful conversations.


\begin{thebibliography}{99}

\bibitem[*]{AJBaddress}
{\em Email:} ajb@maths.uq.edu.au 

\bibitem
{weyl}
Weyl, H., 
Zeitschr. Phys. {\bf 46}, 1--46 (1927); ``The theory of
groups and quantum mechanics," (Dover, New York, 1931), p. 274.

\bibitem{vonneumann}
von Neumann, J.,  
 Math. Ann. {\bf 104}, 570--578 (1931).  

\bibitem{wigner}
Wigner, E.P., 
Phys. Rev. {\bf 40}, 749--759 (1932); 
 Z. Phys. Chem. {\bf B19},
203--216 (1932); 
Lec. Notes Phys.
{\bf 278}, 162--170 (1987); also in W.
Yourgrau and A. van der Merwe (Eds), {\em Perspectives in Quantum
Theory} (MIT Press, Cambridge, Mass., 1971), pp. 26--36.

\bibitem
{groenewold}
Groenewold, H., 
Physica {\bf 12}, 405--460 (1946).

\bibitem
{moyal}
Moyal, J.E., 
Proc. Camb. Phil. Soc. {\bf 45}, 99--124 (1949).

\bibitem{takabayasi}
Takabayasi, T., 
Prog. Theor. Phys. {\bf 11}, 341--373 (1954).  


\bibitem
{stratonovich}
Stratonovich, R.L., 
Sov. Phys. JETP {\bf 4}, 891--898 (1957).

\bibitem{baker}
Baker, G.A. Jr., 
Phys. Rev. {\bf 109}, 2198--2206 (1958).


\bibitem{fairlie}
Fairlie, D.B., 
Proc. Camb. Phil. Soc. {\bf 60}, 581--586 (1964).


\bibitem
{agarwal}
Agarwal, G.S. and Wolf, E., 
Phys. Rev. D {\bf 2}, 2161--2186; 2187--2205; 2206--2225 (1970).


\bibitem
{berezin}
Berezin, F.A., 
Math. USSR Izvestija {\bf 8}, 1109--1165 (1974); 
Commun. Math. Phys. {\bf 40}, 153--174 (1975);
Berezin, F. A. and Subin, M. A., 
in B. Sz.-Nagy (Ed.) {\em Hilbert space operators and operator algebras.
Colloquia Mathematicae Societatis Janos Bolyai 5.} (North Holland,
Amsterdam, 1972),
pp. 21--52.

\bibitem{pool}
Pool, J.C.T., 
J. Math. Phys. {\bf 7}, 66--76 (1966).

\bibitem{berry}
Berry, M.V., 
Philos. Trans. R. Soc. Lond. A {\bf 287}, 237--271 (1977). 


\bibitem 
{bayen} 
Bayen, F., Flato, M., Fronsdal, C., Lichnerowicz, A. and Sternheimer, D., 
Ann. Phys. (N.Y.) {\bf 111}, 
61-110; 111--151 (1978). 

\bibitem{shirokov}
Shirokov, Yu. M., 
Sov. J. Nucl. {\bf 10}, 1--18 (1979). 

\bibitem{berezin3}
Berezin, F.A., 
Sov. Phys. Usp. {\bf 23}, 763--788 (1980). 

\bibitem{tatarskii}
Tatarskii, V.I., 
Sov. Phys. Usp. {\bf 26}, 311--327 (1983).

\bibitem{hillery}
M. Hillery, R.F. O'Connell, M.O. Scully and 
E.P. Wigner, 
Phys. Reps {\bf 106}, 121--167 (1984).

\bibitem{gadella}
Gadella, M., 
Fortschr. Phys. {\bf 43}, 229--264 (1995). 

\bibitem{schroeck}
F.E. Schroeck, {\em Quantum mechanics on phase space} (Kluwer, Boston, 1996).

\bibitem{curtright}
Curtright, T., Fairlie, D. and Zachos, C., 
Phys. Rev. D {\bf 58}, 025002 (1998).

\bibitem{ozorio}
Ozorio de Almeida, A.M., 
Phys. Rep. {\bf 295}, 265--342 (1998). 

\bibitem{manko}
Man'ko, V.I. and Vilela Mendes, R.,  
Physica {\bf D145}, 330-348 (2000).

\bibitem{vallejos}
Vallejos, R.O. and Saraceno, M., 
J. Phys. A {\bf 32}, 7273--7286 (1999). 

\bibitem{bracken}
Bracken, A.J., Doebner, H.-D. and Wood, J.G., 
Phys. Rev. Lett. {\bf 83}, 3758--3761 (1999). 

\bibitem{dubin}
Dubin, D.A., Hennings, M.A. and Smith, T.B., {\em Mathematical aspects of Weyl quantization and
phase}, (World Scientific, Singapore, 2000).

\bibitem{zachos} Zachos, C., 
Int. J. Mod. Phys. A{\bf 17}, 297--316   (2002).

\bibitem
{sternheimer}
Dito, G. and Sternheimer, D., 
{\em IRMA Lec.  Math. and Theoret. Phys., 
Vol. 1},(De Gruyter, Berlin, 2002), pp.
9--54.

\bibitem{segal}
Segal, I.E., 
Math. Scand. {\bf 13}, 31--43 (1963).

\bibitem{kostant}
Auslander, L. and Kostant, B., 
Bull. Amer.
Math. Soc. {\bf 73}, 692--695 (1967);
Kostant, B., 
Lec. Notes. Math. {\bf 170}, 87--208 (1970).

\bibitem{kirillov}
Kirillov, A.A., {\em Elements of the theory of group representations} (Springer-Verlag, Berlin,
1976).

\bibitem{fronsdal}
Fronsdal, C., 
Rep. Math. Phys. {\bf 15}, 111--145 (1978).

\bibitem{gilmore}
Gilmore, R., 
Lec. Notes Phys. {\bf 178}, 211-213 (1978). 

\bibitem 
{agarwal2} 
Agarwal, G.S., 
Phys. Rev. 
A {\bf 24}, 2889--2896 (1981). 

\bibitem 
{arnal} 
Cahen, M. and Gutt, S., 
Lett. Math. Phys. {\bf 6}, 395--404 (1982);
Arnal, D., Cahen, M. and Gutt, S., 
Bull. Acad. Royale Belg. {\bf 74}, 
123--141 (1988); 
Arnal, D., Cahen, M. and Gutt, S., 
Bull. Soc. Math. Belg. {\bf 41}, 207--227 (1989). 

\bibitem{graciabondia}
Gracia-Bondia, J.M. and Varilly, J.C.,
J. Phys. A {\bf 21}, L879--883 (1988); 
Varilly, J.C. and Gracia-Bondia, J.M.,
Ann. Phys. (NY) {\bf 190}, 107--148 (1989); 
Carinena, J.F., 
Gracia-Bondia, J.M. and Varilly, J.C.,
J. Phys. A {\bf 23},
901--933 (1990);
Figueroa, H., 
Gracia-Bondia, J.M. and Varilly, J.C.,
J. Math. Phys. {\bf 31}, 2664--2671 (1990);
Gracia-Bondia, J.M. and Varilly, J.C.,
J. Math. Phys. {\bf 36}, 2691--2701  (1995). 

\bibitem{bertrand}
Bertrand, J. and Bertrand, P., 
J. Math. Phys. {\bf 39}, 4071--4090 (1998). 

\bibitem{antonsen}
Antonsen, F., 
Int. J. Theor. Phys. {\bf 37}, 697--757 (1998). 

\bibitem{brif}
Brif, C. and Mann, A., 
Phys. Rev. A {\bf 59}, 971--987 (1999). 

\bibitem{weigert}
Amiet, J-P. and Weigert, S., 
Phys. Rev. A {\bf 63}, 012102 (2001). 

\bibitem{wolf}
Twareque Ali, S., Atakishiyev, N.M., Chumakov, S.M. and Wolf, K.B., 
Ann. Henri Poincar\'e {\bf 1}, 685--714 (2000).

\bibitem
{moshinsky}
Garcia-Calderon, G. and Moshinsky, M., 
J. Phys. A {\bf 13}, L185--188 (1980); Dirl, R., Kasperkovitz, P. and Moshinsky, M.,
J. Phys. A {\bf 21}, 1835--1846 (1988); Moshinsky, M. and Sharma, A.,
Ann. Phys. (NY) {\bf 282}, 138--153 (2000).

\bibitem{dragt}
Hakioglu, T., 
J. Phys. A {\bf 32}, 4111--4130 (1999); 
Hakioglu, T. and Dragt, 
J. Phys. A {\bf 34}, 6603-6615 (2002).

\bibitem{wigner2}
Wigner, E.P., {\em Group theory and its application to the quantum mechanics of atomic spectra}
(Academic Press, New York, 1959).  

\bibitem 
{bargmann} 
Bargmann, V., 
J. Math. Phys. {\bf 5}, 862--868 (1964). 

\bibitem 
{cassinelli} 
Cassinelli, G., de Vito, E., Lahti, P.J. and Levrero, A., 
Rev. Math. Phys. {\bf 9}, 921--941 (1997). 

\bibitem{leonhardt} 
U. Leonhardt, {\em Measuring the quantum state of light} 
(Cambridge University Press, 1997). 

\bibitem{Abram} 
M. Abramowitz and I.A. Stegun, {\em Handbook of mathematical 
functions} (Dover, New York, 1972). 

\bibitem 
{gelfand} 
Gel'fand, I.M., Shilov, G.E. and Vilenkin, N.Y., ``Generalised
Functions,"
Vols 1--5 (Academic Press, New York, 1964--8). 
 
\bibitem 
{roberts} 
Roberts, J.E., 
J. Math. Phys. {\bf 7}, 1097--1104 (1966). 
 
\bibitem 
{antoine} 
Antoine, J.-P., 
J. Math. Phys. {\bf 10}, 53--69, 2276--2290 (1969). 
 
\bibitem
{bohm}
Bohm, A.and Gadella, M., {\em Dirac kets, Gamow vectors, 
and Gel'fand triplets:  the rigged Hilbert space
formulation of quantum mechanics. Lectures in mathematical physics at the University of Texas at
Austin} (Springer-Verlag, Berlin, 1989).

\bibitem{leblond} 
L\'evy-Leblond, J. -M., 
in E.M. Loebl (Ed.), {\em Group Theory and its Applications II.}, pp 221--299 
(Academic Press, New York, 1971).  

\end{thebibliography}
\end{document}